\documentclass[journal,onecolumn,web]{ieeecolor}
\usepackage{generic}
\usepackage{cite}
\usepackage{amsmath,amssymb,amsfonts}
\usepackage{algorithmic}
\usepackage{graphicx}
\usepackage{textcomp}
\usepackage{chngcntr}
\newtheorem{thm}{Theorem}[section]

\newtheorem{lemma}[thm]{Lemma}
\newtheorem{prop}[thm]{Proposition}
\newtheorem{definition}{Definition}[section]{\bf}{\it}

\newtheorem{remark}{Remark}[section]

\newcommand{\dep}{\operatorname{dep}}
\newcommand{\dom}{\operatorname{dom}}
\newcommand*{\QEDB}{\hfill\ensuremath{\square}}%
\usepackage{dblfloatfix}
\def\BibTeX{{\rm B\kern-.05em{\sc i\kern-.025em b}\kern-.08em
    T\kern-.1667em\lower.7ex\hbox{E}\kern-.125emX}}
\begin{document}
\title{\LARGE \bf A Stochastic Binary Vertex-Triggering Resetting Algorithm for Global Synchronization of Pulse-Coupled Oscillators}
\author{Muhammad U. Javed, Jorge I. Poveda, Xudong Chen \thanks{The authors are with the ECEE Department, CU Boulder. Corresponding author: M. U. Javed.  Emails: \texttt{\{muhammad.javed, jorge.poveda, xudong.chen\}@colorado.edu}. The work is supported by NSF
under Grants ECCS-1809315 and CRII: CNS-1947613, and AFOSR under Grant FA9550-20-1-0076.} 
}

\maketitle

\begin{abstract}
In this paper, we propose a novel stochastic binary resetting algorithm for networks of pulse-coupled oscillators (or, simply, agents) to reach global synchronization. The algorithm is simple to state: Every agent in a network oscillates at a common frequency. Upon completing an oscillation, an agent generates a Bernoulli random variable to decide whether it sends pulses to all of its out-neighbours or it stays quite.  Upon receiving a pulse, an agent resets its state by following a  binary phase update rule. We show that such an algorithm can guarantee global synchronization of the agents almost surely as long as the underlying information flow topology is a rooted  directed graph. The proof of the result relies on the use of a stochastic hybrid dynamical system approach. Toward the end of the paper, we present numerical demonstrations for the validity of the result and, also, numerical studies about the times needed to reach synchronization for various information flow topologies.  
\end{abstract}


\section{Introduction}
\label{sec:intro}
In this paper, we consider a network of $N$ pulse-coupled  oscillators (PCOs), characterized by periodic resetting dynamics, sharing information with their neighbors where the neighboring relations are described by a directed graph (digraph).  Each agent has an individual state $\tau_i\in\mathbb{R}$, which evolves according to the following continuous-time dynamics:
\begin{equation}\label{flows_agents}
\tau_i\in[0,1)\implies\dot{\tau}_i=\frac{1}{T},~~\forall~i\in\{1,2,\ldots,N\},
\end{equation}
where $T>0$ is the period of oscillation, and $[0,1)$ is a normalized unit interval. When the state of an agent $i$ finishes an oscillation, i.e. $\tau_i=1$, it will  instantaneously \emph{reset} its individual state back to zero:
\begin{equation}\label{resets_uncoupled}
\tau_i=1~\implies~\tau_i^+=0.
\end{equation}
Simultaneously, the agent sends a pulse, with a certain probability  $p\in (0,1)$, to trigger {\em all} of its (out-)neighbors $j$. 
Each out-neighbor $j$ of agent $i$, upon receiving the pulse, instantaneously updates its  state $\tau_j$ using a set-valued \textit{binary} phase update rule:
\begin{equation}\label{PUR}
\tau_j^+\in \mathcal{R}_j(\tau_j)= \left\{\begin{array}{cl}
\{0\}&~~ \tau_j \in [0, r_j)\\
\{0,1\} &~~ \tau_j=r_j\\
\{1\} &~~ \tau_j \in (r_j,1]\\
\end{array} \right.,
\end{equation}
 \begin{figure}[t]
    \centering
\includegraphics[width=0.4\textwidth]{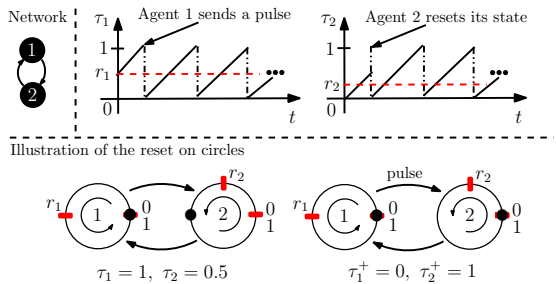} 
   \caption{Illustration of our stochastic binary resetting algorithm: When agent~1 satisfies $\tau_1=1$, it updates its state by $\tau_1^+=0$. Meanwhile, it draws a Bernoulli random variable to decide whether to send a pulse to its out-neighbors (in this case, only agent~2). If the pulse is sent, then upon receiving the pulse, agent~2 will update its state by following the binary update rule described in Eq.~\eqref{PUR}. In this example, since $\tau_2>r_2$, agent 2 update its state by $\tau_2^+=1$. Note that once all agents are synchronized they remain synchronized under our algorithm \eqref{flows_agents}-\eqref{PUR}.
   }
   \label{fig:intro_model}
   \vspace{-0.1cm}
\end{figure}
 where the constant  $r_j\in (0,1)$ partitions the unit interval.

 Amongst others, in this paper we show that if the underlying information flow topology is a rooted directed graph, then for any $p$ and any $r := [r_1,\ldots, r_N]^\top$, the network of PCOs will reach synchronization almost surely.  Since each  individual state $\tau_i$ is confined to evolve in the normalized interval $[0,1]$,  one can view the state as flowing in a unit circle (that is formed by identifying the two endpoints $0$ and $1$ with each other), in the counter-clockwise direction, with frequency $1/T$.  In this way,  global synchronization of PCOs can be cast as a consensus problem on the $N$-torus (e.g. see \cite{ClocksSensitivty,Javed2020ScalableRA,Klinglmayr_2012}). See Figure \ref{fig:intro_model} for an illustration of our algorithm on the $2$-torus.

  \begin{figure*}[t]
    \centering
\includegraphics[width=0.7\textwidth]{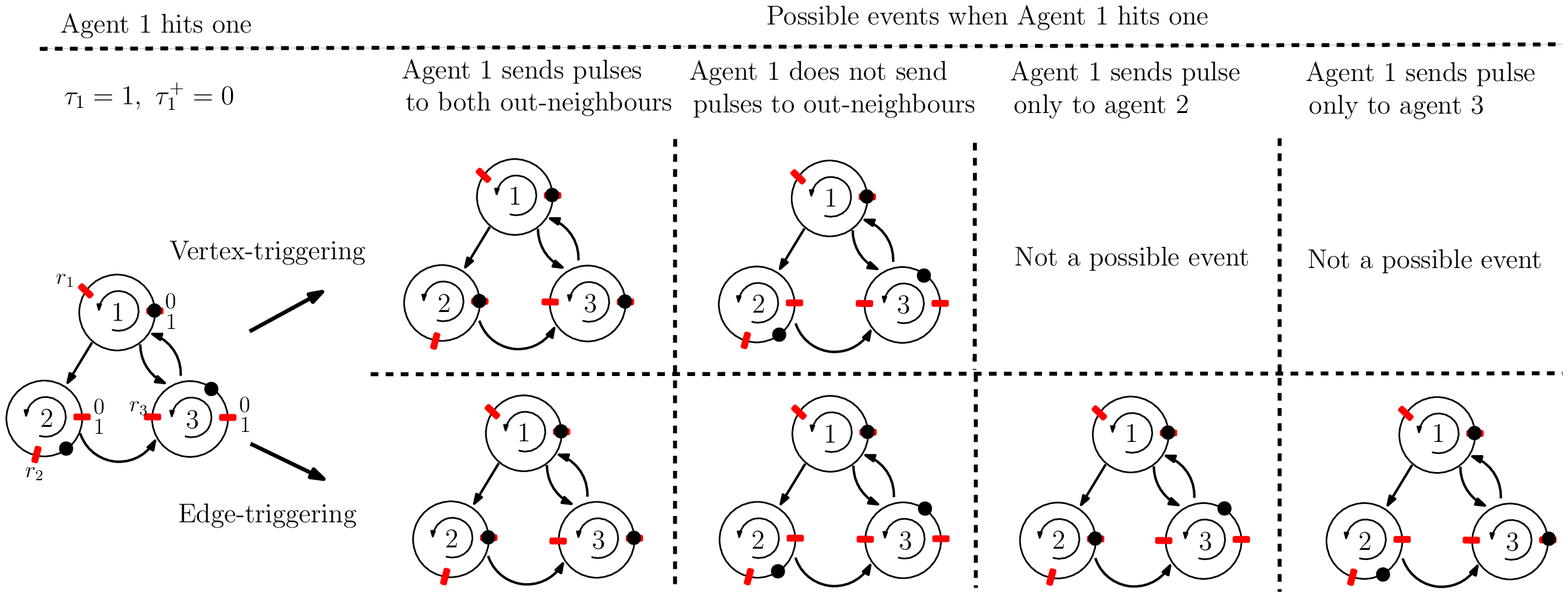} 
   \caption{Two different random triggering models ``vertex-triggering'' and ``edge-triggering.'' We present possible events when agent~1 hits value~1 (i.e., the agent satisfies $\tau_1=1,~\tau_1^+=0$) for both models. An out-neighbour of agent~1 updates its state to either 0 or 1 when it receive a pulse from agent~1; otherwise, the out-neighbor will retain its state.}
   \label{fig:compare_model}
   \vspace{-0.1cm}
\end{figure*}

 Synchronization of PCOs using deterministic resetting algorithms has been widely investigated in the literature, and we refer the reader to \cite{SyncPhillip,cyclic_network,byzantine,Javed2020ScalableRA,NunezSync,GAO2019,DoyleTAC,canavier2017globally,AHDS15,Nishimura,Sepulchre12,Kannapan,wang2020attack,Anton17} and  references therein. However, in none of these works global synchronization is shown to be achieved over all rooted digraphs using deterministic resetting algorithms. Some works have relaxed the global convergence requirement to either \textit{local} convergence (e.g.   \cite{SyncPhillip,Kannapan,wang2020attack}) or \textit{almost global} convergence (e.g. \cite{Nishimura,Sepulchre12,Anton17}) while other works have restrictions on the underlying digraphs~\cite{byzantine,Javed2020ScalableRA,NunezSync,GAO2019,DoyleTAC,canavier2017globally,AHDS15,cyclic_network}.  Recently, we have shown in \cite{Javed2020ScalableRA} that a certain deterministic binary  resetting algorithm cannot achieve global synchronization over all rooted digraphs. Whether or not there exists a deterministic resetting algorithm that can achieve global synchronization of PCOs over all rooted digraphs still remains open.

The problem of global synchronization of PCOs using stochastic resetting algorithms has also been investigated in the literature~\cite{Klinglmayr,Klinglmayr_2012,pagliari,Javed2020ScalableRA,HartmanBook}. Our study on the problem, as well as the main results established in the paper, are different from the ones in those existing works, as we elaborate below. 

First, we mention the references~\cite{Klinglmayr,Klinglmayr_2012}. In these works, the authors have considered a similar stochastic resetting algorithm. A key difference is that their phase update rule is described by a piecewise continuous function (the only discontinuity is at $r_j$, which is set to be 0.5 for all the agents), with each piece being {\em strictly monotonically increasing} whereas ours  is piecewise {\em constant}. Although the difference in the phase update rule seems to be moderate, the analyses of the two resulting systems differ significantly. In particular, the arguments developed in~\cite{Klinglmayr,Klinglmayr_2012} do not apply to our case; certain key results, such as Lemma 8 in \cite{Klinglmayr}, do not hold anymore. For example, the authors there have considered the arc of minimum length that covers all the agents on the unit circle and shown that 
the number of agents on the boundary points of the arc cannot increase over time. This is not true if one uses binary phase update rule. 
Besides the difference in phase update rule, there is also a difference in the underlying network topology.  
Using their resetting algorithm, the authors  
have established almost sure global synchronization over undirected connected graphs (i.e., communications between agents are reciprocal) in~\cite{Klinglmayr_2012} and over strongly connected digraphs in \cite{Klinglmayr}. The class of rooted digraphs considered in this paper is more general.

Next, in the work~\cite{pagliari}, the authors have considered a different type of triggering: Upon hitting~$1$, an agent~$i$ will generate {\em multiple}   independent, identically distributed ({\em i.i.d.}) Bernoulli random variables, with the number of random variables matching the number of its neighbors (that is, the underlying information flow topology is undirected), so as to decide {\em individually} whether or not it sends a pulse to each of its neighbors. This is in contrast to the triggering model considered in this paper where an agent, upon hitting $1$, draws only a {\em single} Bernoulli random variable and {\em broadcasts} to all of its (out-)neighbors. 
Because of this, we call our triggering model {\em vertex-triggering} and theirs {\em edge-triggering}. See Figure \ref{fig:compare_model} for an illustration of both models.
Note that our previous work~\cite{Javed2020ScalableRA} has also considered {\em edge-triggering}. An advantage of ``vertex-triggering'' over ``edge-triggering'' is that the former requires a less number of Bernoulli random variables drawn at a time, making it easier for the agents to implement the resetting algorithm. The difference between the two algorithms will also be carried over to the analysis: For edge-triggering, the underlying network topology can be viewed as an Erd{\H o}s-R\'enyi type random graph whenever an agent hits $1$ (since the edges are drawn independently); in~\cite{Javed2020ScalableRA}, we relied on such a probability model to establish almost sure global synchronization. However, this probability model cannot be used here to describe the network topology for the case of vertex-triggering. Due to the difference between the two probability models, we will have different sample paths of random graphs along the dynamics of the two systems. Consequently, the characterizations of the so-called ``synchronization strings'' (roughly speaking, these are the strings in a sample path that can lead to global synchronization as we will introduce in Definition~\ref{sync_string}) will also be different.

We further mention the work~\cite{HartmanBook} where the authors have considered a completely different stochastic resetting algorithm. There, the dynamics of the agents are not pulse-coupled; instead, the authors have assumed that every agent can access the mean of the states and uses that information to make decisions and to take actions.

Our method to establish almost sure global synchronization relies on the use of stochastic hybrid dynamical systems (SHDS) \cite{sta_rec}, where the set-valued binary update rule will be used to define the jump maps of the system. Indeed, the combination of continuous-time dynamics, describing the continuous evolution of the PCOs, and discrete-time dynamics, describing the resets, naturally lead to a hybrid dynamical system. Moreover, since the pulse-triggering of an agent (upon hitting $1$) is at random and since only the (out-)neighbors of the agent could receive the pulse (if the pulse is generated and sent), the jump maps of the SHDS are stochastic and depend on the underlying information flow topology. Formally, to establish the SHDS, we will first introduce a family of infinite sequences of {\em i.i.d} random digraphs, sampled from a finite set, termed the set of {\em feasible digraphs}. Roughly speaking, a digraph is feasible if every agent is connected to either all or none of its out-neighbors.   
Every such random digraph corresponds to an occurrence of an agent hitting $1$, and it indicates whether the agent sends a pulse or not. We then use such sequence of random digraphs to define the sequence of jump maps of the SHDS. We analyze random solutions of the SHDS by analyzing solutions of a  hybrid dynamical system (HDS) over a fixed, but arbitrary, infinite sequence of feasible digraphs. We present a novel condition on the sequence that can guarantee global synchronization of the HDS. We then establish almost sure global synchronization of the SHDS by showing that the condition can be satisfied almost surely. Toward the end of the paper, we have conducted numerical studies for validation of the main result  and for comparison of our algorithm with an existing vertex-triggering algorithm~\cite{Klinglmayr}.

The remainder of the section is organized as follows: Section \ref{sec:Preliminaries} presents some preliminaries. Main results for the deterministic and stochastic settings are presented and established in Sections \ref{sec:Deterministic} and  \ref{sec:stochastic}, respectively. 

Section \ref{sec:simulation} is about numerical studies. The paper ends with conclusions in Section \ref{sec:conclusion}.  

\vspace{0.1cm}

\noindent
\textbf{Notations.} Given a vector $x$ in $\mathbb{R}^n$,  let $|x|$ be the standard Euclidean norm of $x$.  
For a compact set $\mathcal{A} \subset \mathbb{R}^n$, let $|x|_\mathcal{A} := \min_{y \in \mathcal{A}}|x -y|$.  We also use $|\cdot|$ to denote the cardinality of a finite set. 
We use $\mathbf{c}_n\in\mathbb{R}^n$ to denote a constant vector with all entries equal to $c\in\mathbb{R}$. We use $(a_1,a_2)\succeq (b_1,b_2)$  to denote $a_1 \ge b_1$ and $a_2\ge b_2$. The floor function is denoted by $\lfloor \cdot \rfloor$. Given a set $B$, we use $B^N$ to denote the $N$-Cartesian product of $B$, i.e., $B^N:=B\times B\times\cdots\times B$ ($N$ times). A function $\alpha$ is said to be of class-$\mathcal{K}$ if it is strictly increasing in its argument and $\alpha(0)=0$.  Additionally, if $\alpha(r) \to \infty$ as $r \to \infty$ then $\alpha$  belongs to  class-$\mathcal{K}_\infty$.  
We denote by $\mathbb{B}$ (resp. $\mathbb{B}^\circ$) the closed (resp. open) ball of radius one centered at zero.   A set-valued mapping $M: \mathbb{R}^m \rightrightarrows \mathbb{R}^n$ is said to be locally bounded (LB) at $x \in \mathbb{R}^m$ if there exists a neighborhood $K_x$ of $x$ such that $M(K_x)$ is bounded. Given a set $\mathcal{X} \subset \mathbb{R}^m$, the mapping $M$ is LB relative to $\mathcal{X}$ if the set-valued mapping from $\mathbb{R}^m$ to $\mathbb{R}^n$ defined by $M$ for $x \in \mathcal{X}$,  and by $\varnothing$ for $x \notin \mathcal{X}$, is LB at each $x \in \mathcal{X}$. The graph of a set-valued mapping $G$ is defined as $\text{graph}(G) := \{(x, y) \in  \mathbb{R}^m \times \mathbb{R}^n: y \in G(x)\}$.   Given a measurable space $(\Omega, \mathcal{F})$, a set-valued map $G: \Omega \rightrightarrows \mathbb{R}^n$ is said to be \textit{$\mathcal{F}$-measurable} \cite[Def. 14.1]{Rockafellar}, if for each open set $\mathcal{O} \subset \mathbb{R}^n$, the set $G^{-1}(\mathcal{O}) :=
\{\omega \in \Omega : G(\omega) \cap \mathcal{O} 	= \varnothing\}\in \mathcal{F}$.

\section{Preliminaries}
\label{sec:Preliminaries}
In this section, we present basic notions from graph theory, and deterministic and stochastic hybrid dynamical systems.

\subsection{Graph Theory}

A directed graph, or digraph, is denoted by $\mathcal{G} := (\mathcal{V} , \mathcal{E} )$, with $\mathcal{V} := \{1, 2, \cdots , N\}$ the set of vertices and  $\mathcal{E} \subset \mathcal{V} 
\times \mathcal{V}$ the set of edges. 
In this paper, we consider only simple digraphs, i.e., digraphs without self-arcs. 
We denote by $(i,j)$ an edge of $\mathcal{G}$;   
we call $i$ an in-neighbor of $j$, and $j$ an out-neighbor of $i$. 
We denote the set of  out-edges of vertex $i$ as $\mathcal{E}_i^-$.
A path from a vertex $i$ to a vertex $j$ is a sequence $\{i_0,i_1,\cdots,i_m\}$, with $i_0 = i$ and $i_m = j$, in which each pair $(i_l,i_{l+1}) \in \mathcal{E}$ for all $l \in \{0,1,\cdots,m -1\}$ and all the vertices are pairwise distinct. The length of a path is defined to be the number of edges in that path. A vertex $i \in \mathcal{V}$ is said to be a root of $\mathcal{G}$ if for any other vertex $j \in \mathcal{V}$, there exists a path from $i$ to $j$. A digraph $\mathcal{G}$ with at least one root is a \textit{rooted digraph}. We denote the set of all the root vertices of $\mathcal{G}$ as $\mathcal{V}_R$.   In a rooted digraph $\mathcal{G}$, the \textit{depth of a vertex} $j$ with respect to a given root vertex $i^*$ is defined to be the length of the minimum path from $i^*$ to $j$. We denote by $\mathcal{V}_q(i^*)$ the vertices at depth $q$, and $q^*$ the maximum depth. The \textit{depth of a rooted digraph} $\mathcal{G}$ is defined to be $\dep(\mathcal{G}):=\max_{i^*\in \mathcal{V}_R} q^*$. 
A \textit{$d$-regular digraph} $\mathcal{G}(\mathcal{V},\mathcal{E}')$, for $d<N$, is a digraph where each vertex~$i$ has $d$ out-neighbours $((i + j) \mod N) + 1$, for $j = 0,\ldots,d-1$. Note, in particular, $1$- and $(N-1)$-regular digraphs are {\em cycle} and {\em complete} digraphs, respectively.

\subsection{Hybrid Dynamical System with Random Inputs}

A stochastic hybrid dynamical system (SHDS) with state $x\in\mathbb{R}^n$ and random input $v\in\mathbb{R}^m$ is characterized by the following set of equations \cite{rec_principle,sta_rec}:
\begin{subequations}\label{SHDS1}
	\begin{align}
	&x\in C,~~~~~~~~~\dot{x}= f(x),\label{SHDS_flows0}\\
	&x\in D,~~~~~~x^+\in G(x,v^+),~~~v\sim \mu(\cdot)~~\label{SHDS_jumps0}
	\end{align}
\end{subequations}
where the function $f:\mathbb{R}^n\to\mathbb{R}^n$, called the {\em flow map}, describes the continuous-time dynamics of the system; the set $C\subset\mathbb{R}^n$, called the {\em flow set}, describes the points in the space where $x$ is allowed to evolve according to the differential equation \eqref{SHDS_flows0}; $G:\mathbb{R}^n\times\mathbb{R}^m\rightrightarrows\mathbb{R}^n$, called the {\em jump map}, is a set-valued mapping that characterizes the discrete-time dynamics of the system; and $D\subset\mathbb{R}^n$, called the {\em jump set}, describes the points in the space where $x$ is allowed to evolve according to the stochastic difference inclusion \eqref{SHDS_jumps0}. We use $v^+$ as a place holder for a sequence of independent, identically distributed ({\em i.i.d.}) input random variables $\{{\bf v_k}\}_{k=1}^{\infty}$ with probability distribution $\mu$, derived from an abstract probability space $(\Omega,\mathcal{F},\mathbb{P})$.  

\vspace{0.1cm}
 \begin{definition}\label{definitionbasic1}
A SHDS \eqref{SHDS1} is said to satisfy the \textbf{Basic Conditions} if the following holds: (a) The sets $C$ and $D$ are closed,  $C\subset \text{dom}(f)$, and $D\subset\text{dom}(G)$. (b) The function $f$ is continuous. (c) The set-valued mapping $G:\mathbb{R}^n\times\mathbb{R}^m\rightrightarrows\mathbb{R}^n$ is LB and the mapping $v\mapsto \text{graph}(G(\cdot,v)):=\{(x,y)\in\mathbb{R}^n\times\mathbb{R}^n:y\in G(x,v)\}$ is measurable with closed values. \QEDB
\end{definition}

\vspace{0.1cm}
For further details of SHDS \eqref{SHDS1} (concept of solution, causality assumption, etc), we refer the reader to Appendix \ref{App:SHDS}.

When the discrete-time dynamics \eqref{SHDS_jumps0} does not depend on random inputs, the SHDS \eqref{SHDS1} is reduced to a  standard  hybrid dynamical system (HDS) \cite{teel}: 
\begin{subequations}\label{HDS}
	\begin{align}
	&x\in C,~~~~~~~~~\dot{x}= f(x),\label{HDS_flows1}\\
	&x\in D,~~~~~~~~x^+\in G(x).~~\label{HDS_jumps1}
	\end{align}
\end{subequations}
Solutions of hybrid systems are parameterized by both continuous- and discrete-time indices $t\in\mathbb{R}_{\geq0}$ and $k\in\mathbb{Z}_{\geq0}$. The index $t$ increases continuously during flows  \eqref{SHDS_flows0} or \eqref{HDS_flows1}, and the index $k$ increases by one when a jump occurs via \eqref{SHDS_jumps0} or \eqref{HDS_jumps1}.   Solutions to \eqref{HDS}  are defined on hybrid time domains which are characterized by a pair of time indices $(t,k)$. For further details on the concept of solution to \eqref{HDS} and hybrid time domains, we refer the reader to Appendix \ref{App:HDS}.
 A solution is said to be: (a) \textit{maximal} if its time domain is not a proper subset of the domain of other solution; (b) \textit{complete} if its time domain is unbounded; (c) textit{uniformly non-Zeno} if there exist $\tilde{T},\tilde{K} \in \mathbb{R}_{> 0}$  such that for every $(t_1,k_1),(t_2,k_2) \in \dom(x)$, $t_2-t_1\leq \tilde{T}$ implies that $k_2-k_1\leq \tilde{K}$.

\subsection{Stability and Convergence Notions}

In this paper, we will consider the following properties 
for the solutions of the SHDS \eqref{SHDS1} and the HDS \eqref{HDS}:

\vspace{0.1cm}
\begin{definition}
The HDS \eqref{HDS} is said to render a closed set $\mathcal{A}$ \textbf{strongly forward invariant} \cite{teel}  if every complete solution $x$, with $x(0,0)\in \mathcal{A}$, satisfies  $x(t,k) \in \mathcal{A}$ for all $(t,k) \in \dom(x)$. Similarly, the  SHDS \eqref{SHDS1} is said to render the set $\mathcal{A}$  \textbf{surely strongly forward invariant} if every random solution $x_{\omega}$ to \eqref{SHDS1} with $x_{\omega}(0,0)\in\mathcal{A}$ stays surely in $\mathcal{A}$. \QEDB
\end{definition}
\vspace{0.1cm}

Next, we introduce some stability and convergence notions from \cite{teel, sanfelice2020hybrid}:

\vspace{0.1cm}

\begin{definition}\label{UGFTS}
The HDS~\eqref{HDS}  renders a closed set $\mathcal{A}$

\begin{enumerate}
    \item Uniformly Globally Stable (\textbf{UGS})  if there exists a class-$\mathcal{K}_\infty$ function $\alpha$ such that any solution $x$ to \eqref{HDS} satisfies $|x(t,k)|_{\mathcal{A}} \leq \alpha(|x(0,0)|_{\mathcal{A}})$  for all $(t,k) \in \dom (x);$
    \item Globally Finite-Time Attractive (\textbf{GFTA})  if for each solution $x$ of \eqref{HDS} there exists $\bar{T}(x(0,0))>0$ such that $|x(t,k)|_\mathcal{A} = 0$ for all $(t,k)\in \dom (x)$ and $t+k\ge \bar{T}(x(0,0))$.
    \item Globally Fixed-Time Attractive (\textbf{GFxTA}) if $\mathcal{A}$ is GFTA and, additionally, $\bar{T}>0$ is a constant independent of $x(0,0)$.  \QEDB
\end{enumerate}
 
\end{definition}

\vspace{0.1cm}

We further have the following definition from \cite{sta_rec}, which applies to systems of the form \eqref{SHDS1}:

\vspace{0.1cm}
\begin{definition}\label{SHDS2def}
The SHDS \eqref{SHDS1}  renders a compact set $\mathcal{A}$:
\begin{enumerate}
\item \textbf{Uniformly Lyapunov stable in probability}  if for each $\varepsilon>0$ and $\rho>0$ there exists a $\delta>0$ such that for all $\mathbf{x}_\omega(0,0) \in \mathcal{A}+\delta \mathbb{B}$, every maximal random solution ${\bf x}_\omega$ from $\mathbf{x}_\omega(0,0)$ satisfies the inequality:

\vspace{-0.4cm}
\begin{small}
\begin{align}\label{UGASpDef}
&\mathbb{P}\Big(\mathbf{x}_{\omega}(t,k)\in\mathcal{A}+\varepsilon\mathbb{B}^\circ,~\forall~ (t,k)\in\dom(\mathbf{x}_{\omega})\Big)\geq 1-\rho.
\end{align}
\end{small}

\vspace{-0.4cm}
\item \textbf{Uniformly Lagrange stable in probability} if for each $\delta>0$ and $\rho >0$, there exists $\varepsilon>0$ such that the inequality \eqref{UGASpDef} holds. 
\item \textbf{Uniformly globally attractive in probability} if for each $\varepsilon>0, \rho>0$ and $R>0$, there exists $\gamma\geq 0$ such that for all random solutions  ${\bf x}_\omega$ with $\mathbf{x}_\omega(0,0)\in \mathcal{A}+R \mathbb{B}$ the following holds: 
\begin{align*}
\mathbb{P}\Big(&\mathbf{x}_{\omega}(t,k)\in\mathcal{A}+\varepsilon\mathbb{B}^\circ,\forall~t+k\geq \gamma,(t,k)\in \dom(\mathbf{x}_{\omega})\Big)\notag \\
&~~~~~~~~~~~~~~~~~~~~~~~\geq 1-\rho.
\end{align*}
\end{enumerate}
System \eqref{SHDS1} is said to render a compact set $\mathcal{A}\subset\mathbb{R}^n$ \textbf{Uniformly Globally Asymptotically Stable in Probability (\textbf{UGASp})} if it satisfies conditions (a), (b), and (c). \QEDB
\end{definition}

For a SHDS of the form \eqref{SHDS1}, UGASp of a compact set can  be established via the stochastic hybrid invariance principle \cite[Thm. 8]{rec_principle}, see Theorem \ref{S_rec}.

\section{Deterministic Resetting Algorithm with Time-varying Jump Maps}
\label{sec:Deterministic}

In this section, we first introduce a deterministic HDS, with time-varying jump maps, for analyzing the asymptotic behavior of a typical random solution of the networked system of PCOs described in Section~\ref{sec:intro}. The results of this section will be used later  to establish almost sure global synchronization in Section~\ref{sec:stochastic}.

\subsection{Well-posed Hybrid Dynamical System  }

To formalize the  HDS, we start by introducing a notion about feasible subgraphs of a given digraph: 

\vspace{0.1cm}
\begin{definition}
	Let $\mathcal{G} = (\mathcal{V}, \mathcal{E})$ be an arbitrary digraph. A subgraph $\phi = (\mathcal{V}, \mathcal{E}')$, on the same vertex set $\mathcal{V}$, is \textbf{feasible} if the edge set $\mathcal{E}'$ satisfies the following condition: For any vertex $i\in \mathcal{V}$, either $\mathcal{E}^-_i\subseteq \mathcal{E}'$ or $\mathcal{E}^-_i\cap \mathcal{E}' = \varnothing$. \QEDB
\end{definition}

Let  $\Phi$ be the collection of feasible subgraphs of $\mathcal{G}$ and $\Xi$ be the collection of infinite sequences of the feasible subgraphs, i.e., any element $\xi \in \Xi$ is given by $\xi:=\phi_1\phi_2\phi_3\cdots$, where each $\phi_{i} \in \Phi$ is feasible. 
We note that both $\Phi$ and $\Xi$ are implicitly dependent of $\mathcal{G}$. 
Now, for a given $\xi \in \Xi$, we define a corresponding HDS:  
	\begin{equation}\label{eq:HDS}
	\mathcal{H}_\xi:=(C,f,D,G).
	\end{equation} 
	The state of the HDS is $x:=(\tau,\lambda)\in \mathbb{R}_{\ge 0}^N \times\mathbb{Z}_{\geq0}$. 
	The continuous-time dynamics of this system are given by
	\begin{equation}\label{continuoushybrid1}
	\dot{x}=f(x):=\begin{bmatrix}
f_\tau(\tau)\\f_\lambda(\lambda)
	\end{bmatrix}=\begin{bmatrix}
\frac{1}{T}\mathbf{1}_N\\0
	\end{bmatrix},
	\end{equation}
	where the state $x$ evolves in the set $C$ defined as,
	\begin{equation}\label{flowset}
C:=C_\tau \times\mathbb{Z}_{\geq0},~~~C_\tau:=[0,1]^N.
	\end{equation}
	The discrete-time dynamics are given by
	\begin{equation}\label{jump_map_deterministic_in}
	x^+ \in G(x):= \begin{bmatrix}
	G_{\lambda+1}(\tau)\\ \lambda+1
	\end{bmatrix},
	\end{equation}
where the state $x$ evolves in the set $D$ defined as,
	\begin{equation}\label{jumpset}
	D:= D_\tau \times\mathbb{Z}_{\geq0}, ~~D_\tau:=\big\{\tau\in C_\tau:\text{max}_{i\in\mathcal{V}}~\tau_i=1\big\}.
	\end{equation}
	Note that in \eqref{continuoushybrid1}-\eqref{jumpset}, the sub-state $\lambda$ can be viewed as a discrete-time counter that increases by one every time there is a jump in the system. For each $\lambda$, the set-valued map $G_\lambda$ in \eqref{jump_map_deterministic_in} is defined as the outer-semicontinuous hull of the mapping
	\begin{align}\label{mappings}
	&G_{\lambda}^0(\tau)=\left\{g\in \mathbb{R}^N: g_i=0,~g_j\in \mathcal{R}_{j,\lambda}(\tau),\; \forall j \neq i \right\},
	\end{align}\noindent 
where the set-valued map $\mathcal{R}_{j,\lambda}$ is defined over the feasible digraph $\phi_\lambda:=(\mathcal{V},\mathcal{E}_\lambda)$ as: 
\begin{equation*}
\tau_j^+ \in \mathcal{R}_{j,\lambda}(\tau) :=\left\{\begin{array}{cll}
0 & \tau_j \in [0, r_j)& (i,j) \in \mathcal{E}_\lambda,\\
\{0,1\} & \tau_j=r_j& (i,j) \in \mathcal{E}_\lambda,\\
1 & \tau_j \in (r_j,1]& (i,j) \in \mathcal{E}_\lambda,\\
\tau_j & & (i,j) \not\in \mathcal{E}_\lambda.\\
\end{array} \right.
\end{equation*} 


For any $r\in(0,1)^N$, and for any $\xi \in \Xi$, every solution of the HDS~\eqref{eq:HDS} is complete and uniformly non-Zeno. The completeness follows from the fact that the HDS~\eqref{eq:HDS} satisfies the Basic Conditions of Definition \ref{definitionbasic1} and the following facts: (i) $f_\tau(\tau)>0$ for all $x\in C\setminus D$, which guarantees existence of non-trivial solutions from $C\setminus D$; (ii) the system has no finite escape times; and (iii) $G(D)\subset C \cup D$, so solutions cannot stop due to jumps leaving $C\cup D$, \cite[Prop. 6.10]{teel}. 
The property that~\eqref{eq:HDS} is uniformly non-Zeno follows from the next result:



\vspace{0.1cm}
\begin{lemma}\label{lem:correspondence}
	Consider a HDS $\mathcal{H}_\xi$ as in \eqref{eq:HDS}.  Let $\underline{r}:=\min_{i\in\mathcal{V}}r_i$. Then, the number of jumps in any period of length $T$ is bounded below and above  by $1$ and $N (\lfloor 1/\underline{r} \rfloor + 1)$, respectively.  \QEDB
\end{lemma}
\vspace{.1cm}

\begin{proof} We first establish the lower bound. Pick an arbitrary agent $i$ and let $\tau_i(t,k)$ be its state at time $(t,k)$. We consider the period $[t, t+T]$. There are two cases: (1) During this period, no in-neighbor of agent $i$ hits $1$ and triggers it. In this case, by the continuous-time dynamics of the HDS~\eqref{continuoushybrid1}, agent $i$ will reach value $1$ in at most $T$ seconds and, then, jumps to $0$; (2) During that period, there exists at least one in-neighbor of agent $i$ that hits $1$ and jumps. In either case, the total number of jumps of the entire network that occur during the period $[t, t+ T]$ is bounded below by $1$.    
	
	We now establish the upper bound. For each individual agent, we will evaluate an upper bound for the number of times it can hit $1$. 
	To do so, we note that if an agent $i$ hits $1$ and jumps at a certain time $(t, k)$ (so that $\tau_i(t,k+1) = 0$), then for the next $r_iT$ seconds, the agent cannot hit $1$. This holds because the least time for the agent $i$ to hit value $1$ is to first flow for $r_i T$ seconds and, then, to have one of its in-neighbors to hit $1$ and trigger it.   
	The above arguments then show that the number of times the agent $i$ can hit $1$ during the period $[t, t+T]$ is bounded above by $(1/r_i + 1)$ and, hence, 
	$\lfloor 1/r_i \rfloor + 1$. Finally, because the number of jumps of the entire network that occur during the period $[t, t + T]$ is equal to the number of times the agents hit $1$ during the same period, we conclude that the number of jumps is bounded above by
	$
	\sum^N_{i = 1} \left ( \lfloor 1/r_i  \rfloor + 1 \right ) \le N (\left \lfloor 1/{\underline{r}} \right \rfloor + 1).
	$
\end{proof} 
\subsection{Stability Analysis}

We study the stability properties of the HDS, introduced in \eqref{eq:HDS}, with respect to the closed set $\mathcal{A}$ defined as follows: 
\begin{equation}\label{synchronization_set}
	\mathcal{A}:=\mathcal{A}_s \times \mathbb{Z}_{\ge 0},~\mathcal{A}_s:= \{\mu \mathbf{1}_N: \mu \in [0,1]\} \cup \{0,1\}^N.\end{equation}
	
It should be clear that $x\in \mathcal{A}$ if and only if $\tau\in \mathcal{A}_s$. Note that if the  sub-state $\tau$ reaches the compact set $\mathcal{A}_s$, then the network reaches synchronization. To proceed, we first have the following result, which says that the network remains synchronized once it achieves synchronization: 
\vspace{.2cm}	
	
\begin{lemma} \label{forward_invariant}
For any $\xi\in \Xi$, the HDS $\mathcal{H}_\xi$, introduced in~\eqref{eq:HDS}, renders the set $\mathcal{A}_s$ (resp. $\mathcal{A}$) strongly forward-invariant for the sub-state $\tau$ (resp. the state $x$). \hfill $\blacksquare$
\end{lemma}
\vspace{.2cm}

 We provide a proof of the above lemma in Appendix~\ref{App:proof}.

Recall that for a root~$i^*$ of $\mathcal{G}$, the set of vertices at depth $q$ is denoted by $\mathcal{V}_q(i^*)$. 
We will now introduce the notion of  {\em synchronization string}: 

\vspace{.1cm}
\begin{definition} \label{sync_string}
	Let $\mathcal{G}=(\mathcal{V},\mathcal{E})$ be a rooted digraph and $i^* \in \mathcal{V}_R$ be a root.
	Let $q^*$ be the depth of $\mathcal{G}$ with respect to~$i^*$.
	For any $q = 0,\ldots, q^*$, we 
	let $\mathcal{G}_q:= (\mathcal{V}, \mathcal{E}_q)$ be a feasible subgraph of $\mathcal{G}$ with the edge set
	$
	\mathcal{E}_q:= \cup_{i\in \mathcal{V}_q(i^*)} \mathcal{E}^-_i.
	$
	Then,
	the \textbf{{synchronization string}} $\zeta$ with respect to~$i^*$ is a finite string of feasible subgraphs of $\mathcal{G}$:
	\begin{equation}\label{stringofgoodgraphs}
	\zeta := \mathcal{G}_0\cdots\mathcal{G}_0\mathcal{G}_1 \cdots \mathcal{G}_1 \cdots  \mathcal{G}_{q^*-1},\cdots\mathcal{G}_{q^*-1},
	\end{equation}
	where each subgraph $\mathcal{G}_q$ is repeated contiguously in the string for  $\ell^*$ times, where
$
	    \ell^*:=N (\lfloor 1/\underline{r} \rfloor + 1).
$
 Correspondingly, the length  of the string $\zeta$ is $L^*:=\ell^*q^*$.
\end{definition}
\begin{figure*}
    \centering
    \includegraphics[width=0.7\textwidth]{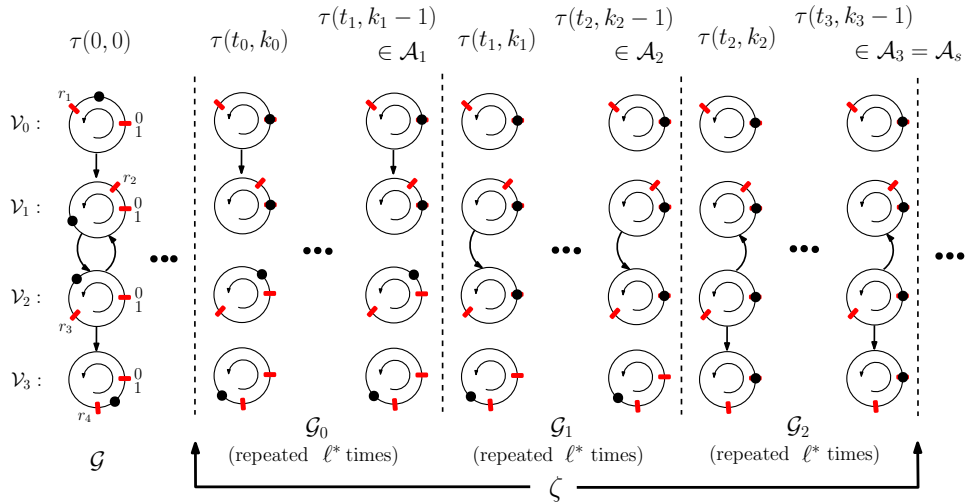}
    \caption{To illustrate Theorem \ref{thm:deterministic}, we consider a network  of four PCOs of depth $q^*=3$ with vertices $\mathcal{V}_0=\{1\}, \mathcal{V}_1=\{2\}, \mathcal{V}_2=\{3\}, \mathcal{V}_3=\{4\}$, edge set $\{(1,2),(2,3),(3,2),(3,4)\}$ and $\underline{r}=1/8$. Correspondingly, $\ell^*=36$ and the length of the synchronization string $\zeta$ becomes $108$. For the initial condition $\tau(0,0)$ and infinite sequence of graphs $\xi$, $(t_{q-1},k_{q-1})$ and $(t_{q},k_{q}-1)$, for $q=1,2,3$, corresponds to the first and last indices respectively of $\mathcal{G}_q$ (in $\zeta$)  in  $\xi$.  Since, $\mathcal{G}_q$ only contains the out-edges of vertices $\mathcal{V}_q$, this leads to  $\tau(t_q,k_q-1)\in \mathcal{A}_q$ for every $q$. Because, for such a $\xi$, each $\mathcal{A}_q$ is strongly forward-invariant, we have that $\tau(t_3,k_3-1)\in \mathcal{A}_s$.   }
    \label{fig:deter_result1}
\end{figure*}
With the definition above, we will now state the first main result of the paper: 

\vspace{.2cm}

\begin{thm}\label{thm:deterministic}
	Let $\mathcal{G}$ be a rooted digraph and $r \in (0, 1)^N$. Suppose that  $\xi\in \Xi$ contains a synchronization string $\zeta$, defined in \eqref{stringofgoodgraphs}, with respect to a root $i^* \in \mathcal{V}_R$; 
	then,  for every maximal solution $x$ of the corresponding HDS  $\mathcal{H}_\xi$, defined in \eqref{eq:HDS}, with $\lambda(0,0)=0$ and $\tau(0,0)\in [0,1]^N$, there exists a hybrid time $(t^*,k^*)\in \dom(x)$,  uniformly bounded above,  such that $\tau(t,k) \in \mathcal{A}_s$ for all $(t,k)\succeq (t^*,k^*) $. 
\end{thm}

\vspace{.2cm}

We establish below Theorem~\ref{thm:deterministic}. 
To proceed, we will first introduce a few subsets $\mathcal{A}_q(i^*)$, for $q = 0,\ldots, q^*$, that describe partial synchronizations in the network. For each $q$, let $\tau^{(q)}\in [0,1]^{|\mathcal{V}_q(i^*)|}$ be the vector that collects the states of  vertices of depth $q$, with respect to $i^*$ (so that $\tau^{(0)}=\tau_{i^*}$). Next, we relabel (if necessary) the vertices in the rooted digraph $\mathcal{G}$ so that:
\begin{equation}
    \tau=[\tau^{(0)};\tau^{(1)};\cdots;\tau^{(q^*)}].
    \label{tau_label}
\end{equation} 
In the sequel, we fix the root $i^*$ and the corresponding labelling.  
Let $\mathcal{U}_q(i^*):= \cup^q_{\ell = 0} \mathcal{V}_{\ell}(i^*)$
and $M_q:=|\mathcal{U}_q(i^*)|$. Then, the subsets  $\mathcal{A}_q(i^*)$  are defined as follows:
\begin{equation}\label{synchronization_subset}
	\mathcal{A}_{q}(i^*):= \biggl(\{\mu \mathbf{1}_{M_q}: \mu \in [0,1]\} \cup \{0,1\}^{M_q}\biggr) \times [0,1]^{N-M_q}.
\end{equation}
Note, in particular, that if $q = 0$, then $\mathcal{A}_{0}(i^*)=C_\tau$ and if $q = q^*$, then $\mathcal{A}_{q^*}(i^*)=\mathcal{A}_s$, where $\mathcal{A}_s$ is defined in \eqref{synchronization_set}. It should be clear that if the sub-state $\tau$ belongs to one of these compact sets, say $\mathcal{A}_q(i^*)$ at a certain time, then the vertices in $\mathcal{U}_q(i^*)$ are synchronized at that time. By definition, we have the following chain of inclusions:
\begin{equation} \label{filtration}
   \mathcal{A}_{q^*}(i^*) \subsetneq \mathcal{A}_{q^*-1}(i^*)\subsetneq\cdots \subsetneq \mathcal{A}_1(i^*) \subsetneq  \mathcal{A}_0(i^*). 
\end{equation}
%
%

Let $\xi =  \phi_1\phi_2\cdots$ be given in the statement of Theorem \ref{thm:deterministic}. 
Because $\xi$ contains the synchronization string $\zeta$, there exists a $k_0\in \mathbb{Z}_{\geq 1}$  such that $\zeta = \phi_{k_0} \cdots  \phi_{{k_{q^*}}-1}$, where $k_{q^*}$ is such that $(k_{q^*} - k_0)$ is the length of $\zeta$. 
Further, for each $q = 1,\ldots, q^*-1$, we let 
\begin{equation}\label{eq:defk0}
    k_q:=k_0+ q\ell^*,
\end{equation}
where $q^*$ is the depth of $\mathcal{G}$ with respect to the root~$i^*$ and $\ell^*$ is defined in Definition \ref{sync_string}. It should be clear that $k_{q^*}$ can be expressed as $k_{q^*} = k_0 + q^*\ell^*$. 
From the definition of synchronization string, we have that for any $q = 0,\ldots, q^*-1$, the digraphs $\phi_{k_q},\ldots, \phi_{k_{q+1}-1}$ are the same given by $\mathcal{G}_q$. 

Now, let $x$ be an arbitrary maximal solution of the HDS $\mathcal{H}_\xi$ and we fix this solution in the sequel. 	
Since $\lambda(0,0)=0$, we have that $\lambda(\cdot,k)=k$, for all $(t,k)\in\dom(x)$.

Let  $t_0:=\min\{t\in\mathbb{R}_{\geq0}:(t,k_0)\in\dom(x)\}$, i.e., $t_0$ is the continuous time-instant that corresponds to  the occurrence of the $k_0$th jump of the HDS $\mathcal{H}_\xi$. %
Note that such $t_0$ is well-defined  and, in fact, uniformly bounded above. Indeed, to see this, note that by Lemma \ref{lem:correspondence}, the number of jumps in any period of $T$ is lower bounded by $1$. It thus implies that $t_0\le k_0 T$. Similarly, for any other $q = 1,\ldots,q^*$, we 
let 
\begin{equation}
   t_{q}:= \min\{t\in\mathbb{R}_{\geq0}:(t,k_q)\in\dom(x)\}.
   \label{t_k}
\end{equation}
Again, from Lemma \ref{lem:correspondence}, there is a  uniform upper bound for $t_q$ as $t_{q}\le t_0 + q\ell^* T$. See Figure~\ref{fig:deter_result1} for illustration of $(t_q,k_q)$.

With the sets $\mathcal{A}_q(i^*)$ and the hybrid times $(t_q,k_q)$ defined above, we establish the following result: 
\vspace{0.2cm}
\begin{prop}\label{prop:induction}
Let $x$ be a maximal solution of the HDS $\mathcal{H}_\xi$~\eqref{eq:HDS} and $(t_q,k_q)$, for $q = 0,\ldots, q^*$, be the hybrid times defined as above. Then, under the assumption of Theorem~\ref{thm:deterministic}, the following holds: 
 For each $q = 0,\ldots,q^*-1$, there exists a hybrid time $(t'_q,k'_q)\in \dom(x)$, with $(t_q,k_q)\preceq (t'_q,k'_q) \preceq (t_{q+1},k_{q+1}-1)$, such that $\tau(t'_q,k'_q)\in \mathcal{A}_{q+1}(i^*)$, where $\mathcal{A}_{q+1}(i^*)$ is defined in~\eqref{synchronization_subset}.  
 Moreover, $\tau(t,k)\in \mathcal{A}_{q+1}(i^*)$ for all $(t,k)\in \dom(x) $ with 
    $
    (t,k)\succeq (t'_q,k'_q). 
    $
\end{prop}
\vspace{0.2cm}

\noindent
\begin{proof}
We will show that there exist 
hybrid times $(t'_q,k'_q)$, for $q = 0,\ldots, q^*-1$, with $(t_q,k_q)\preceq (t'_q,k'_q) \preceq (t_{q+1},k_{q+1}-1)$, such that 
if $(t'_{q},k'_{q})\preceq (t,k) \preceq (t'_{q+1},k'_{q+1})$, for $q = 0,\ldots, q^*-2$, then 
$\tau(t,k)\in \mathcal{A}_{q+1}(i^*)$. Moreover, $\tau(t'_{q^*-1}, k'_{q^*-1}) \in \mathcal{A}_{q^*}(i^*) = \mathcal{A}_s$. Note that if this holds, then by  chain of inclusion~\eqref{filtration} and the strong forward invariance of $\mathcal{A}_s$ established in Lemma~\ref{forward_invariant}, we have that for any $(t,k)\succeq (t'_q,k'_q)$, $\tau(t,k) \in \mathcal{A}_{q+1}$ for all $q = 0,\ldots, q^*-1$.

Starting from the hybrid time $(t_0,k_0)$, all elements in the string $\phi_{k_0}\cdots \phi_{k_1-1}$ are the same given by the digraph $\mathcal{G}_0$, which induce the set-valued mappings $G_{k_0},\cdots, G_{k_1-1}$ defined in  \eqref{mappings}.  
		    Since the root vertex $i^*$ has no in-neighbor in $\mathcal{G}_0$, 
		    for any state $\tau(t_0,k_0)$, $i^*$ will hit $1$ in less than or equal to $T$ seconds (in continuous-time) after $t_0$, i.e., there exists a hybrid time $(t'_0, k'_0-1)$ such that $\tau^{(0)}(t'_0, k'_0-1) = 1$ with $0\leq  t_0' - t_0 \leq T$ and $\tau^{(0)}(t'_0, k'_0) = 0$. Furthermore, by Lemma~\ref{lem:correspondence}, the number of jumps over any period of length $T$ is bounded above by~$\ell^*$ (defined in Definition \ref{sync_string}), we have that $k_0\leq  k'_0\leq k_{1} - 1$. Because $\mathcal{G}_0$ contains only the out-edges of the root vertex $i^*$, each of the set-valued mapping $G_{k_0}, \cdots, G_{k_1-1}$ maps $\tau(t'_0, k'_0)$ to $\{0,1\}^{M_1}\times [0,1]^{N-M_1}$, where we recall that $M_1$ is the cardinality of the set $\mathcal{U}_1(i^*) = \{i^*\} \cup \mathcal{V}_1(i^*)$.  
			Thus,  we have that $\tau(t'_0,k'_0)\in \mathcal{A}_{1}(i^*)$.

Since $\tau(t'_0,k'_0)\in \mathcal{A}_{1}(i^*)$ and since  $\phi_{k'_0+1},\ldots,\phi_{k_1 -1}$ contain only the out-edges of the root~$i^*$, we have that 
\begin{multline}\label{eq:proof1}
 \tau(t,k)\in \mathcal{A}_{1}(i^*), \mbox{for all $(t,k)\in\dom(\tau)$ such that} \\ (t'_0,k'_0) \preceq (t,k) \preceq (t_{1},k_{1}-1).
 \end{multline}
Starting from the hybrid time $(t_1,k_1)$, all  elements in the string  $\phi_{k_1}\cdots \phi_{k_2-1}$ are the same given by the digraph  $\mathcal{G}_1$, which induce the set-valued mappings $G_{k_1},\cdots, G_{k_2-1}$ defined in~\eqref{mappings}. 
By construction of $\mathcal{G}_1$, only vertices in $\mathcal{V}_1(i^*)$ have out-neighbors. Thus, if there is a jump at a hybrid time $(t,k)$, with $(t_1,k_1)\preceq (t,k) \preceq (t_2,k_2-1)$, then only the vertices in $\mathcal{V}_1(i^*)$ can trigger. 
On one hand, by Lemma~\ref{lem:correspondence}, the number of jumps that can occur during any period of length $T$ is bounded above by $\ell^* = k_2 - k_1$. 
On the other hand, since the vertices in $\mathcal{U}_1(i^*)$ are synchronized at $(t_1,k_1-1)$, starting from $t_1$ these vertices will hit $1$ simultaneously (in continuous-time) 
in less than or equal to $T$ seconds.   The above arguments imply that there exists a hybrid time $(t'_1,k'_1)$, with $t_1 \leq t'_1 \leq t'_1 + T$ and $k_1 \leq k'_1 \leq k_2 - 1$, such that $[\tau^{(0)},\tau^{(1)}](t'_1,k'_1) = \mathbf{0}_{M_1}$. 
Furthermore, each of the set-valued mappings $G_{k_1}, \cdots, G_{k_2-1}$ maps $\tau(t'_1, k'_1)$ to $\{0,1\}^{M_2}\times [0,1]^{N-M_2}$, where $M_2$ is the cardinality of the set $\mathcal{U}_2(i^*) = \{i^*\} \cup \mathcal{V}_1(i^*) \cup \mathcal{V}_2(i^*)$.  
Thus, we have that $\tau(t'_1,k'_1)\in \mathcal{A}_2(i^*)$.  
Note that by the above arguments, we have also shown that $\tau(t,k)\in \mathcal{A}_1(i^*)$ for $(t'_0,k'_0)\preceq (t,k) \preceq (t'_1,k'_1)$. 

One can iterate the above arguments to obtain sequentially the hybrid times $(t'_q,k'_q)$, for all $q = 0,\ldots, q^*-1$, as described in the beginning of the proof. \end{proof}

\vspace{0.2cm}
Theorem~\ref{thm:deterministic} follows immediately from   Proposition~\ref{prop:induction}:

\vspace{0.3cm}

\textit{Proof of Theorem~\ref{thm:deterministic}:} 
Let $x = (\tau, \lambda)$ be a maximal solution. Then, by Proposition~\ref{prop:induction}, there is a hybrid time $(t'_{q^*-1},k'_{q^*-1})$ such that $\tau(t, k)\in \mathcal{A}_{s}$ for all $(t,k)\succeq (t'_{q^*-1},k'_{q^*-1})$. We then set $(t^*,k^*):= (t'_{q^*-1},k'_{q^*-1})$. Further, since $(t^*,k^*) \preceq (t_{q^*},k_{q^*})$ and since $(t_{q^*},k_{q^*})$ is uniformly bounded above, $(t^*,k^*)$ is uniformly bounded above as well.  
\hfill $\blacksquare$

\vspace{.2cm}

Toward the end of the section, we consider the scenario where $\xi$ contains the synchronization string infinitely often. Precisely, we have the following definition: 

\vspace{.2cm}

\begin{definition}
\label{uniform_inf_often}
 An infinite sequence $\xi\in \Xi$ contains a synchronization string {\em infinitely often} if it has infinitely many disjoint finite strings that are the synchronization string.  The infinite sequence $\xi$ contains a synchronization string {\em uniformly infinitely often} if there exists a positive integer~$n$ such that every string of length $n$ in $\xi$ contains a synchronization string. 
\end{definition}
%

\vspace{.2cm}

With Definition~\ref{uniform_inf_often}, we will now strengthen Theorem~\ref{thm:deterministic}:
\vspace{0.2cm}
\begin{thm}
\label{thm:UGAS}
Let $\mathcal{G}$ be a rooted digraph and $r\in (0,1)^N$. 
Let the HDS $\mathcal{H}_\xi$ be given as in \eqref{eq:HDS} and the set $\mathcal{A}$ be defined in~\eqref{synchronization_set}.  Suppose that $\xi$ contains the synchronization string $\zeta$ infinitely often (resp. uniformly infinitely often);   
    then, $\mathcal{H}_\xi$ renders the set $\mathcal{A}$ UGS and GFTA (resp. UGS and GFxTA).

\end{thm}
\vspace{0.2cm}

\noindent

\begin{proof} 
We show that $\mathcal{H}_\xi$ renders the set $\mathcal{A}$ uniformly globally stable and  globally  finite-time  attractive (resp. globally  fixed-time  attractive) when $\xi$ contains $\zeta$ infinitely often (resp. uniformly infinitely often):
\vspace{0.1cm}

\noindent
\textit{Proof of  uniform global stability of $\mathcal{A}$:} Consider the  function $V:[0,1]^N\to \mathbb{R}_{\geq0}$ defined as the infimum of all the arcs that cover all agents on the unit circle, where the points $0$ and $1$ are identified to be the same.  The mathematical expression of $V$ can be given as follows~\cite{Klinglmayr}:  
\begin{align}
\label{Lyapunov_function}
   V(\tau):=1-\max_{1\leq i \leq N} \begin{cases}
\tau_{\gamma_{i+1}}-\tau_{\gamma_{i}}&\text{for } i <N,\\
1-\tau_{\gamma_{i}}+\tau_{\gamma_{1}}&\text{for } i =N,
\end{cases} 
\end{align}
where $\gamma_i$, for $i \in \{1,\cdots, N\}$, is an index permutation such that
$\tau_{\gamma_{i}}\leq \tau_{\gamma_{i+1}}$  for all $i$. 
This function satisfies the following properties: (a) It is positive definite with respect to the compact set $\mathcal{A}_s$ defined in \eqref{synchronization_set}; (b) It remains constant during flows because all the oscillators have the same frequency $\frac{1}{T}$; (c) It does not increase at jumps since  jumps never increase the number of distinct positions of the agents.  Next, we  define a Lyapunov  function candidate  $W:[0,1]^N \times \mathbb{Z}_{\ge 0}\to \mathbb{R}_{\geq0}$ for the HDS $\mathcal{H}_\xi$ as follows: for any $x = (\tau,\lambda)$, let 
$W(x):=V(\tau)+|\lambda|_{\mathbb{Z}_{\geq 0}}$. 
 Since $\lambda \in \mathbb{Z}_{\geq 0}$,  we have that $W(x)=V(\tau)$ and $|x|_\mathcal{A}=|\tau|_{\mathcal{A}_s}$. Thus, it suffices to  show that the  HDS $\mathcal{H}_\xi$ renders the set $\mathcal{A}_s$  uniformly globally stable for the sub-state $\tau \in C_\tau$. We establish this fact below. 
 Since $C_\tau$ is compact and since $V$ is continuous, $V$ is bounded; in fact, it is known~\cite{Javed2020ScalableRA} that $V(\tau)\leq 1-\frac{1}{N}$. 
 Because $V$ is positive definite with respect to compact set $\mathcal{A}_s$, there exists two class $\mathcal{K}$ functions $\alpha_1,\alpha_2$ such that the condition $\alpha_1(|\tau|_{\mathcal{A}_s})\leq V(\tau)\le \alpha_2(|\tau|_{\mathcal{A}_s})$ holds for all $\tau \in C_\tau$ \cite[Chp. 3]{teel}. Next, we know from properties (b) and (c) above that $V$ is non-increasing, i.e.  $V(\tau(t,k))\leq V(\tau(0,0))$ for all $(t,k) \in \dom(x)$.
Thus, we have from above inequalities that $|\tau(t,k)|_{\mathcal{A}_s}\leq \alpha_1^{-1}(\alpha_2(|\tau(0,0)|_{\mathcal{A}_s}))$ for all $(t,k) \in \dom(x)$. Since $\alpha_1^{-1}\alpha_2$ is a $\mathcal{K}$-function,  we have that the HDS $\mathcal{H}_\xi$ renders the set $\mathcal{A}_s$  uniformly globally  stable for the sub-state $\tau \in C_\tau$ \cite{teel}.

\vspace{0.1cm}

\noindent
\textit{Proof of  global finite-time attractivity of $\mathcal{A}$:}
Let $x = (\tau,\lambda)$ be an arbitrary maximal solution of $\mathcal{H}_\xi$. We first establish global finite-time attractivity under the assumption that $\xi$ contains the synchronization string $\zeta$ infinitely often. Specifically, we need to show that there exists a  $\bar{T}( x(0,0))$ such that
\begin{equation}
    |x(t,k)|_{\mathcal{A}}=0,~~\forall~ t+k\geq  \bar{T}(x(0,0)),~ (t,k) \in \dom(x).
\end{equation}
Given $\xi$, let  $\sigma_{i}$ be  the index of~$\xi$ corresponding to the first digraph of the $i$th appearance  of~$\zeta$. Since $\xi$ contains $\zeta$ infinitely often,  there exists an integer $i > 0$  such that  $\sigma_i > \lambda(0,0)$ and $\sigma_j \leq \lambda(0,0)$ for all $j < i$. In words, the integer~$i$ is such that the $i$th appearance of $\zeta$ in the sequence $\xi$ is its first appearance after index $\lambda(0,0)$. If, further, $\xi$ contains $\zeta$ uniformly infinitely often, then, by Definition \ref{uniform_inf_often}, $\sigma_i - \lambda(0,0)\le n$. 
Now, let $$\bar k_0:=\sigma_i-\lambda(0,0), ~\bar t_0:=\min\{t\in \mathbb{R}_{\ge 0}: (t,\bar k_0)\in \dom(x)\}.$$ 
Then, $(\bar t_0, \bar k_0)$ is the hybrid time of the solution~$x$ corresponding to the first digraph of the first appearance of~$\zeta$. 
By Lemma~\ref{lem:correspondence}, we have that $\bar t_0\leq \bar k_0 T$. 
Next, similar to $(t_q,k_q)$ defined in~\eqref{eq:defk0} and~\eqref{t_k}, we let
\begin{align*}  \bar k_q := \bar k_0 +q\ell^*, ~\bar t_q :=\min\{t\in\mathbb{R}_{\geq0}:(t,\bar k_q)\in\dom(x)\}.
\end{align*}
 Using again Lemma~\ref{lem:correspondence}, we have that $\bar t_q - \bar t_0\leq q \ell^* T$.
 Thus, all the hybrid times $(\bar t_q, \bar k_q)$ are bounded above. Furthermore, by the same arguments of Proposition \ref{prop:induction}, we have that 
 $|x(t,k)|_{\mathcal{A}}=0$ for all $(t,k)\succeq (\bar t_{q^*}, \bar k_{q^*})$, with $(t,k)\in \dom (x)$.
The proof of global finite-time attractivity is then done by setting 
\begin{equation}\label{eq:barTx}
\bar{T}(x(0,0)):=\bar t_{q^*}+ \bar k_{q^*}.
\end{equation} Finally, we assume that  $\xi$ contains $\zeta$ {\em uniformly} infinitely often and establish global fixed-time attractivity. 
We do so by showing that the quantity $\bar{T}(x(0,0))$ in~\eqref{eq:barTx} is uniformly bounded above. By the above arguments, we have the following two facts: (1) Since $\bar k_0=\sigma_{i} - \lambda(0,0)\leq n$ and $\bar k_{q^*}=\bar k_0+ q^*\ell^*$, we have that $\bar k_{q^*}  \leq n + q^*\ell^*$; (2) Since $\bar t_0\leq \bar k_0 T \leq n T$ and since $\bar t_{q^*} \leq \bar t_0+ q^* \ell^* T$, we have that $\bar t_{q^*}\leq (n + q^*\ell^*)T$. Thus, for any $x(0,0)$, $\bar T(x(0,0))\leq (n + q^*\ell^*)(T +1)$. 
\end{proof}

\section{Stochastic Resetting Algorithm}
\label{sec:stochastic}

In this section, we consider networks of PCOs  with the underlying information flow topology being a random digraph: Every time an agent hits~$1$, it will generate a {\em single} Bernoulli random variable, independent of others, to decide whether or not to send pulses to {\em all} of its out-neighbors.  This stochastic model differs from the one in our previous work~\cite{Javed2020ScalableRA}; there, whenever an agent hits~$1$, it will generate {\em multiple}  {\em i.i.d.} Bernoulli random one for {\em each} of its  out-neighbors.

\subsection{Well-Posed Stochastic Hybrid Dynamical System}

 We start by showing that the feasible subgraphs of a given digraph $\mathcal{G}$ can be mapped one-to-one to certain binary sequences.   
To that end, let $N^*\le N$ be the number  of vertices in $\mathcal{G}$ with at least one out-neighbor. Without loss of generality, we will label these vertices as $1,\ldots, N^*$ and let $\mathcal{V}^*:=\{1,\ldots, N^*\}$.  
Consider binary sequences $v_1\ldots v_{N^*}$ of length $N^*$, with each $v_i\in \{0,1\}$. 
One can assign to each feasible digraph $\mathcal{G}' = (\mathcal{V}, \mathcal{E}')$ such a binary sequence: For each $i = 1,\ldots, N^*$, set
$$
v_i := 
\begin{cases}
1 & \mbox{if } \mathcal{E}_i \subseteq \mathcal{E}', \\
0 & \mbox{otherwise}. 
\end{cases}
$$
Conversely, each binary sequence gives rise to a feasible digraph. Thus, with the labeling of the vertices in $\mathcal{V}^*$ and the above correspondence, we can use a binary sequence $v_1 \ldots v_{N^*}$ to represent a feasible digraph.  Consequently, the set $\Phi$ can be realized as the collection of all binary sequences of length $N^*$, denoted as $\Psi:=\{0,1\}^{N^*}$.

We next introduce a simple  model that can generate a random feasible digraph.   Let the digits $v_1,\ldots, v_{N^*}$ of a binary sequence $v$ be {\em i.i.d.} Bernoulli $(p)$ random variables, i.e., the probability that $v_i$ takes value 1 (resp. 0) is $p$ (resp. $(1 - p)$). We denote by $\mu$ the corresponding probability measure on $\Phi$.  
It follows that for any feasible digraph $\phi \in \Phi$ represented by a binary sequence $v:=v_1\ldots v_{N^*}$,
\begin{equation}
  \mu(\phi)=p^{N'}(1-p)^{N^*-N'},
  \label{random_graph_model}
\end{equation}
where $N'$ is the total number of $1$'s in the binary sequence. 

With the above random model, we can now construct a stochastic hybrid dynamical system (SHDS).  
First, we consider set-valued mappings $S_{ij}:[0,1]\times \Psi\rightrightarrows[0,1]$, defined for each edge $(i,j) \in \mathcal{E}$ of $\mathcal{G}$ as follows:
\begin{equation}\label{pre_jump_map_stochastic}
S_{ij}(\tau_j,v)=v_i\bar{\mathcal{R}}_j(\tau_j)+(1-v_i)\tau_j,
\end{equation}
where $\bar{\mathcal{R}}_j$ is the outer semi-continuous hull of mapping $\mathcal{R}_j$, defined in \eqref{PUR},
and $v_{i}$ is the digit in a binary sequence that corresponds to the vertex $i$ in $\mathcal{V}^*$.  
Next, using \eqref{pre_jump_map_stochastic}, we define a new set-valued mapping $G^0_S:[0,1]^N\times \Psi\rightrightarrows\mathbb{R}^N$ as follows: 
\begin{align}\label{initial_stochastic_jump_map}
\Bigg\{g\in\mathbb{R}^N:g_i=0,
 g_j\in \left\{\begin{array}{ll} S_{ij}(\tau_j,v), & (i,j)\in\mathcal{E}\\
\{\tau_j\}, &  (i,j)\notin\mathcal{E}
\end{array}\right\}\Bigg\},
\end{align}
where $g_j$ is defined for all $j\neq i$ and the mapping $G_S^0(\tau,v)$ is nonempty only when $\tau_i=1$ for some $i\in\mathcal{V}$ and $\tau_j\in[0,1)$ for $j\neq i$. We used  the subindex $S$ to indicate that the mapping $G^0_S$ is stochastic. Finally, the jump map for the SHDS is defined as the outer-semicontinuous hull of $G_S^0$, i.e., 
\begin{equation}\label{set_valued_jump_map}
G_S(\tau,v):=\overline{G^0_S(\tau,v)}.
\end{equation}
Note that when a jump occurs and a random digraph $\phi_k\in \Phi$ is drawn, not every edge of $\phi_k$ plays a role in the jump map $G_S$. Only the edges $(i,j)$ with $\tau_i = 1$ for some $i\in \mathcal{V}$, matter. 

Let $\omega := \omega_1\omega_2\omega_3\cdots$ be a sequence of {\em i.i.d.} random variables, with each $\omega_i \sim \mu(\cdot )$ a feasible digraph. 
We denote by $\Omega$ the collection of sample paths $\omega$. 
 Each sample path $\omega$ will be used to determine the sequence of jump maps at all discrete times through~\eqref{set_valued_jump_map}.

It follows that the resulting SHDS depends on three parameters, namely,  the parameter $p$ associated with the Bernoulli random variable, the partition vector $r$, and the digraph $\mathcal{G}$.
We will thus write the SHDS as 
\begin{equation}\label{SHDS_graph}
\mathcal{H}_S(p, r, \mathcal{G}):=(C_\tau,f_\tau,D_\tau,G_S),
\end{equation}
where $f_\tau,C_\tau,D_\tau$ are defined in \eqref{continuoushybrid1},  \eqref{flowset}, \eqref{jumpset} and again the subindex $S$ indicates that the overall system is stochastic.  

Note that the HDS in \eqref{eq:HDS}, parameterized by an infinite string of feasible digraphs $\xi$,  is the deterministic counterpart of the SHDS \eqref{SHDS_graph} where the sample path $\omega$ is realized as $\xi$.  It should be clear from our probability model that $\omega$ contains the synchronization string $\zeta$ infinitely often almost surely. 

To proceed, we first have the following fact:  

\vspace{.2cm}
\begin{lemma}
 For any $p\in (0,1)$, any $r\in(0,1)^N$, and any digraph $\mathcal{G}$, 
the SHDS $\mathcal{H}_S(p,r,\mathcal{G})$ satisfies the basic conditions. Moreover, every maximal random solution of $\mathcal{H}_S$ is surely complete and uniformly non-Zeno.
\label{lem:well_posed} \hfill $\blacksquare$
\end{lemma}

\vspace{.2cm}
The proof of the lemma relies on Lemma \ref{lem:correspondence} and uses similar arguments to the ones in the proof of Lemma 5 in~\cite{Javed2020ScalableRA}. We thus omit it here.

\subsection{Global Synchronization with Probability One}

We start by noting the following fact: For any $p\in (0,1)$, $r\in (0,1)^N$, and any digraph $\mathcal{G}$, the SHDS $\mathcal{H}_S(p,r,\mathcal{G})$ renders the  set $\mathcal{A}_s$ surely strongly forward invariant. We omit the proof of such a fact as it uses the same arguments as the ones in the proof of Lemma~\ref{forward_invariant}.

\begin{figure*}
    \centering
   \includegraphics[width=0.7\textwidth]{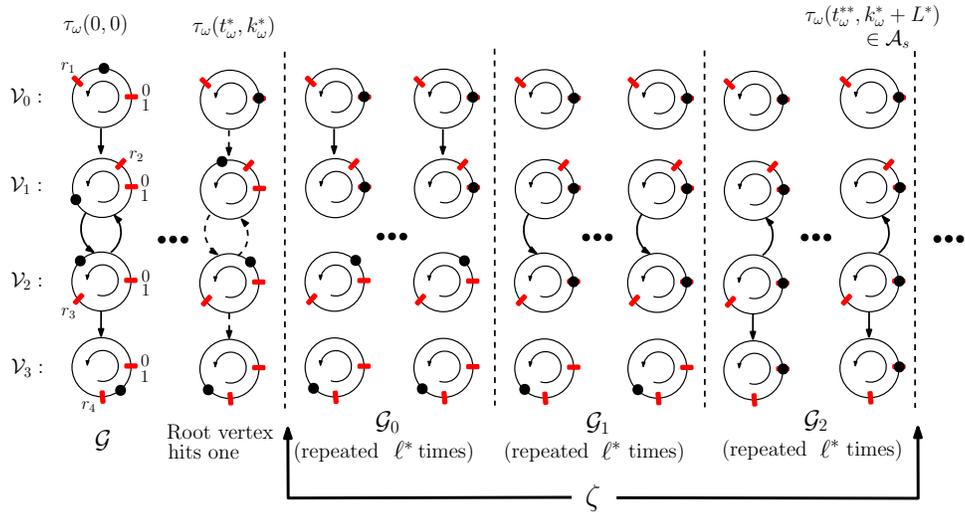}
    \caption{To illustrate Theorem \ref{thm:main}, we consider the same network of PCOs from  Figure \ref{fig:deter_result1}.  For the initial condition $\tau_\omega(0,0)$, the root vertex hits 1 at $(t_\omega^*,k_\omega^*)$ where $t_\omega\leq T$. Followed by that, the synchronization string $\zeta$ appears in $\omega$ from indices $k_\omega^*+1$ to $k_\omega^*+L^*$ where  $L^*=108$. This leads to $\tau_\omega(t_\omega^{**},k_\omega^*+L^*)\in \mathcal{A}_s$ where intermediate steps are shown in Figure \ref{fig:deter_result1}.}
    \label{fig:result_stochastic}
\end{figure*}

Next, similar to our earlier work in \cite{Javed2020ScalableRA},  we introduce the following definition of the {\em sync-triplet} for the SHDS $\mathcal{H}_S$.

\vspace{.2cm}

\begin{definition}\label{compatibility2}
Let $\mathcal{A}_s$ be given in~\eqref{synchronization_set}. Let $p\in (0,1)$, $r\in (0,1)^N$, and $\mathcal{G}$ be a digraph of $N$ vertices. Then,  $(p, r, \mathcal{G})$ is a \textbf{sync-triplet} if the following two items hold: 
\begin{enumerate}
\item For every initial condition in $C\cup D$, there exist  non-trivial random solutions almost surely, and every maximal random solution of $\mathcal{H}_{S}(p,r,\mathcal{G})$ is complete and uniformly Non-Zeno almost surely;
\item The SHDS $\mathcal{H}_S(p,r,\mathcal{G})$ renders $\mathcal{A}_s$  UGASp (see Def.~\ref{SHDS2def}).
\end{enumerate}
\end{definition}

\vspace{.2cm}

A random solution of $\mathcal{H}_S(p, r, \mathcal{G})$ depends on $\omega$ and we denote it by $\pmb{\tau}_\omega$. Note that there may exist multiple random solutions even if we fix $\omega$ and the initial condition, which is due to the set-valued nature of the jump map in the SHDS \eqref{SHDS_graph}. For each random solution $\pmb{\tau}_\omega$, we define:
\begin{equation}\label{firsthitting}
\pmb{T^*}(\pmb{\tau}_{\omega})=\inf\left\{t \mid \pmb{\tau}_{\omega}(t_\omega,k_\omega)\in\mathcal{A}_s,~(t_\omega,k_\omega)\in\dom(\pmb{\tau}_{\omega})\right\},
\end{equation}
which is the (continuous) time that the random solution $\pmb{\tau}_{\omega}$ enters the compact set $\mathcal{A}_s$ defined in \eqref{synchronization_set}.  We call $\pmb{T^*}(\pmb{\tau}_{\omega})$ the \textbf{sync-time} of the random solution.
\vspace{0.1cm}

Now, we will present the main result of this section: 
 
\vspace{.2cm}
\begin{thm}
\label{thm:main}
For any $p \in (0,1)$, any $r\in (0,1)^N$, and any rooted digraph $\mathcal{G}$, $(p, r, \mathcal{G})$ is a sync-triplet. Moreover, for any initial condition $\pmb{\tau}_\omega (0,0)$, the following holds for all positive integers $n$ and all random solutions  $\pmb{\tau}_\omega$ of the SHDS:
\begin{equation}\label{stochasticbound}\mathbb{P}(\pmb{T}^*( \pmb{\tau}_\omega) > n T^*) \leq  \rho^n,\end{equation} where  $T^*:=(\dep(\mathcal{G})\ell^*+1)T$, with $\ell^*$ defined in Definition \ref{sync_string}, and $\rho \in (0, 1)$ is a constant given
by:
\begin{align}\label{eq:definerho}
\rho:=1-(p~(1-p)^{\dep(\mathcal{G}) -1})^{ N\ell^*}.
\end{align}
\end{thm}

\vspace{0.3cm}

\begin{remark}\normalfont
Theorem \ref{thm:main} can further be generalized to the case where agents have heterogeneous probabilities $p_i \in (0,1)$. Correspondingly, the  constant  $\rho$ on the right hand side of~\eqref{stochasticbound} changes to   $1-(\underline{p}~(1-\overline{p})^{\dep(\mathcal{G})-1})^{N\ell^*}$ where $\underline{p}:=\min_{i \in \mathcal{V}^*} p_i$ and $\overline{p}:=\max_{i \in \mathcal{V}^*} p_i$. With slight modification, the arguments below can be used to establish the heterogeneous case.
\end{remark}

 Before presenting the proof of Theorem~\ref{thm:main}, we need a few preliminary results. First, we let $\mathcal{S}(\pmb{\tau}_\omega(0,0))$ be the set of all maximal random solutions of \eqref{SHDS_graph} from the initial condition $\pmb{\tau}_{\omega}(0,0) \in [0,1]^N$.
For each  initial condition, we define the following event: 
\begin{align}
    &\Omega_1(\pmb{\tau}_{\omega}(0,0)):= \Big\{\omega \in \Omega \mid  \forall~\pmb{\tau}_{\omega}\in\mathcal{S}(\pmb{\tau}_{\omega}(0,0)),\exists~ i^*\in \mathcal{V}_R~,\notag\\&
    \exists~(t^*_\omega, k^*_\omega)\in\dom(\pmb{\tau}_{\omega})~\mbox{with $t^*_\omega\le T$}\notag
    ~~\mbox{s.t.}~\pmb{\tau}_{\omega, i^*}(t^*_\omega, k^*_\omega) = 1\Big\}. 
\end{align}
 In words, the above event is about having a certain root vetex~$i^*$ in the network hitting~$1$ before continuous-time~$T$.
Similar to~\cite[Lemma 6]{Javed2020ScalableRA}, the following result holds: 
\vspace{0.2cm}
\begin{lemma}\label{lemmagraph2}
For any $ \pmb{\tau}_\omega(0,0) \in [0,1]^N, \Omega_1(\pmb{\tau}_\omega(0,0)) = \Omega$. 
\end{lemma}
\vspace{0.2cm}
For a positive integer $\ell$  and  a  root vertex  $i^*$ of $\mathcal{G}$, we define an event $\Omega_2(\ell,  i^*)$, by using the synchronization string $\zeta$  from Definition \ref{sync_string}, as follows: 
\begin{equation}\label{eq:defineOmega2}
\Omega_2(\ell,  i^*):= \{\omega\in \Omega \mid \omega_{\ell+1}\cdots \omega_{\ell+{L^*}} =  \zeta\},
\end{equation}  
where $L^*$ is the length of the string $\zeta$. 
We compute below the probability of this event.
\vspace{0.2cm}
\begin{lemma}\label{lem:probability}
Let $\ell^*$ be given in Definition \ref{sync_string}. Then, 
	\begin{equation}\label{eq:exponentialconvergence}
	\mathbb{P}(\Omega_2(\ell,i^*))\geq  (p~(1-p)^{\dep(\mathcal{G}) -1})^{ N\ell^*}.
	\end{equation}
\end{lemma}
\vspace{0.3cm}
\begin{proof}  Recall from Definition \ref{sync_string} that each digraph $\mathcal{G}_q$ in $\zeta$    is a subgraph of the rooted digraph $\mathcal{G}$ with the same vertex set but contains only the out-edges of the vertices at depth $k$ with respect to the root vertex $i^*$. Also, each $\mathcal{G}_q$ is a feasible digraph and it follows from \eqref{random_graph_model}  that $$ \mu(\mathcal{G}_q)=p^{|\mathcal{V}_q(i^*)|}(1-p)^{N^*-|\mathcal{V}_q(i^*)|},$$
where we recall that $\mathcal{V}_q(i^*)$ is the set of vertices at depth $q$ with respect to the root vertex $i^*$. Using  the fact that the random variables $\omega_q$, for $q \ge 1$, are {\em i.i.d.}, we evaluate the probability of the event $\Omega_2(\ell,i^*)$ as follows:
\begin{align*}
\mathbb{P}(\Omega_2(\ell,i^*)) &=\prod_{q=0}^{q^*-1}(p^{|\mathcal{V}_q(i^*)|}(1-p)^{N^*-|\mathcal{V}_q(i^*)|})^{\ell^*}, \\
& = (p^{\sum_{q=0}^{q^*-1}|\mathcal{V}_q(i^*)|} (1-p)^{N^*q^* - \sum_{q=0}^{q^*-1}|\mathcal{V}_q(i^*)|})^{\ell^*}, \\
& = ( p^{N^*} (1-p)^{N^*q^* - N^*} )^{\ell^*}, \\
& \geq ( p (1-p)^{\dep(\mathcal{G}) - 1} )^{N\ell^*},
\end{align*}
where the third equality follows from the fact that  $\sum_{q=0}^{q^*-1}|\mathcal{V}_q(i^*)|=N^*$ and the last inequality follows from the fact that $N^* \leq N$ and $q^* \leq \dep(\mathcal{G})$.
\end{proof} 

\vspace{0.2cm}

With the above preliminary results, we prove  Theorem~\ref{thm:main}:

\vspace{0.2cm}

\textit{Proof of Theorem~\ref{thm:main}:}
We again consider  the function $V:[0,1]^N\to\mathbb{R}_{\geq0}$ as introduced in \eqref{Lyapunov_function}, i.e.,  the infimum of all arcs that cover all agents on the unit circle. 
Using three properties described in the proof of Theorem~\ref{thm:UGAS}, we have that $V$ (positive definite w.r.t. $\mathcal{A}_s$) is non-increasing on average along the random solutions of~\eqref{SHDS_graph} and, hence,  serves as a valid Lyapunov function for the SHDS  (c.f. Appendix \ref{App:SHDS}).


By Lemma \ref{lem:well_posed}, the SHDS \eqref{SHDS_graph} satisfies the basic conditions and  every maximal random solution $\pmb{\tau}_\omega$ of the SHDS is surely complete and uniformly non-Zeno.  Thus, by the stochastic hybrid invariance principle (c.f. Theorem \ref{S_rec}), in order to show that the set $\mathcal{A}_s$ is UGASp, it suffices to show that there does not exist a complete random solution  $\pmb{\tau}_{\omega}$ that remains in a non-zero level set of the Lyapunov function almost surely.

To establish the above fact, we will show that there exist positive constants $\eta$ and $T^*$ such that for any sample path $\omega$ and for any initial condition $\pmb{\tau}_\omega(0, 0)$, the following holds: 
\begin{equation}\label{eq:defineOmega4}
\mathbb{P}(\Omega_3(\pmb{\tau}_\omega(0,0)))> \eta,
\end{equation}
where the event $\Omega_3(\pmb{\tau}_\omega(0,0))$ is given by
\begin{align*}
&\Omega_3(\pmb{\tau}_\omega(0,0)):= \Big\{\omega \in \Omega \mid~\forall~\pmb{\tau}_{\omega}\in\mathcal{S}(\pmb{\tau}_{\omega}(0,0)),~\forall~t_\omega \ge T^*\\&
~\mbox{s.t.}~(t_\omega,k_\omega)\in\dom(\pmb{\tau}_{\omega}),~V(\pmb{\tau}_\omega(t_\omega,k_\omega))=0\Big\}.
\end{align*}
We show below that $\eta$ and $T^*$ can be chosen to be the following values $\eta:= 1 -\rho$, where $\rho$ is defined in 
\eqref{eq:definerho}, and $T^* = (\dep(\mathcal{G})\ell^*+1)  T$.
\begin{figure*}
    \centering
    \includegraphics[width=0.25\textwidth]{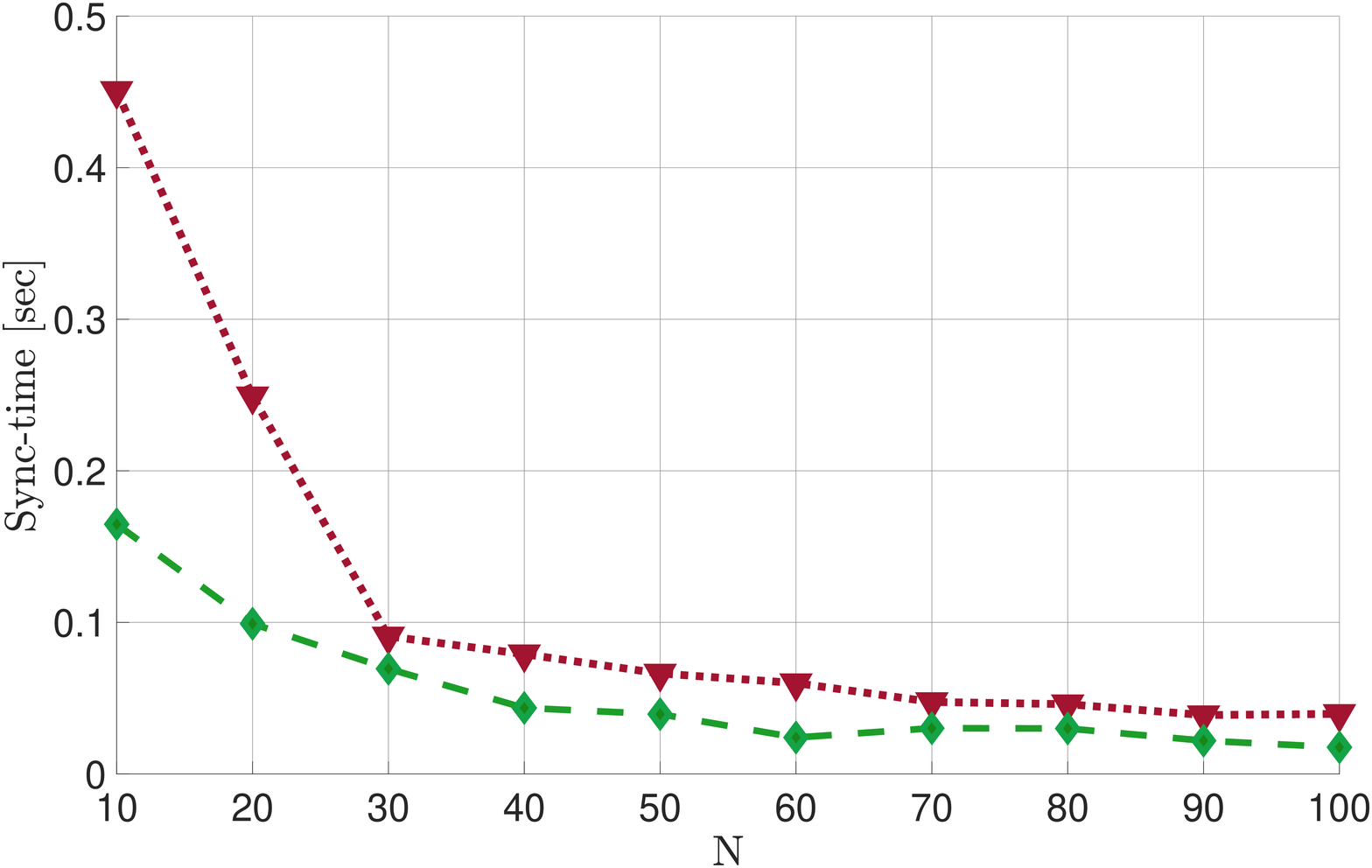}
    \hspace{-0.3cm}  
   \includegraphics[width=0.25\textwidth]{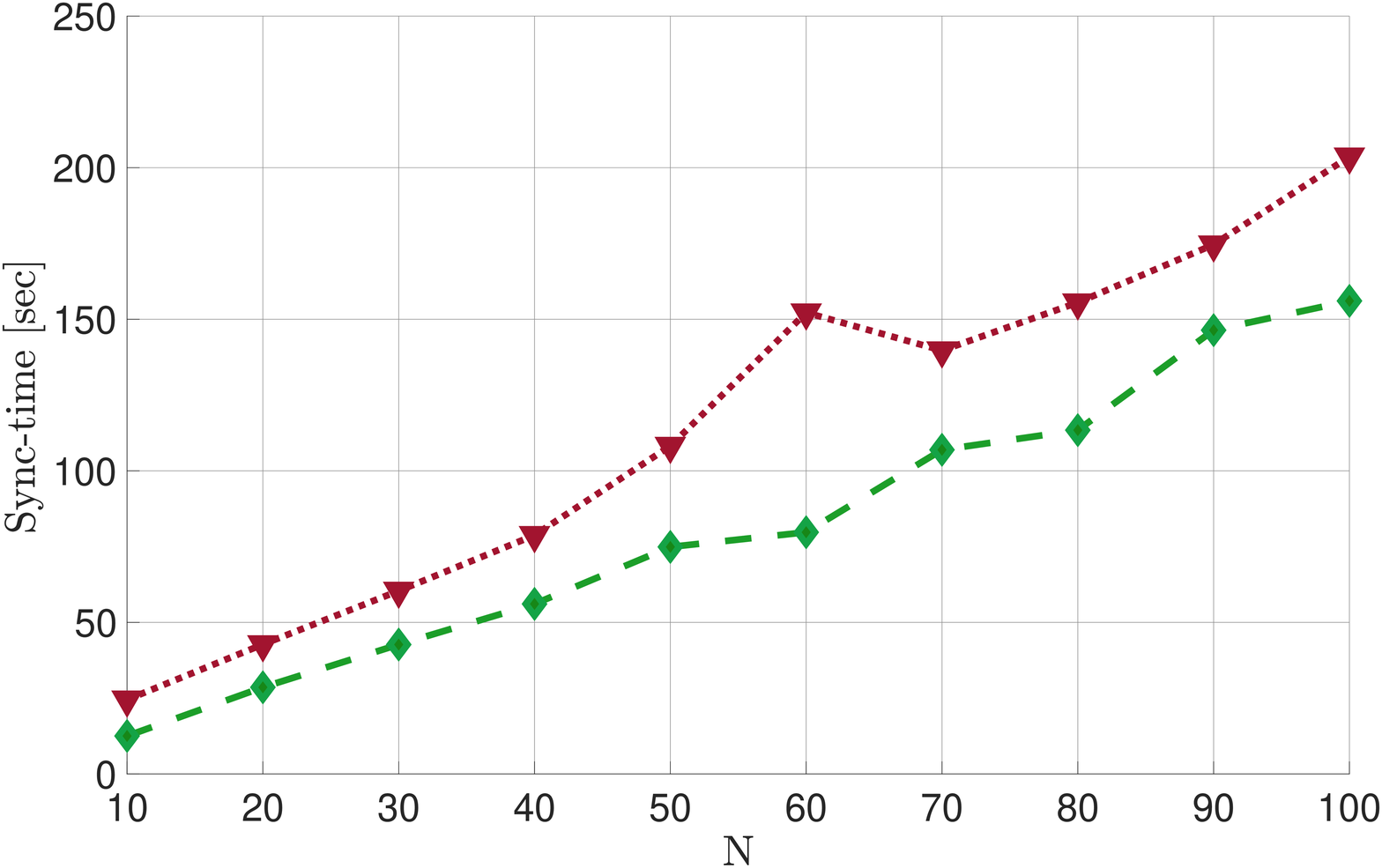}
    \hspace{-0.3cm}  \includegraphics[width=0.25\textwidth]{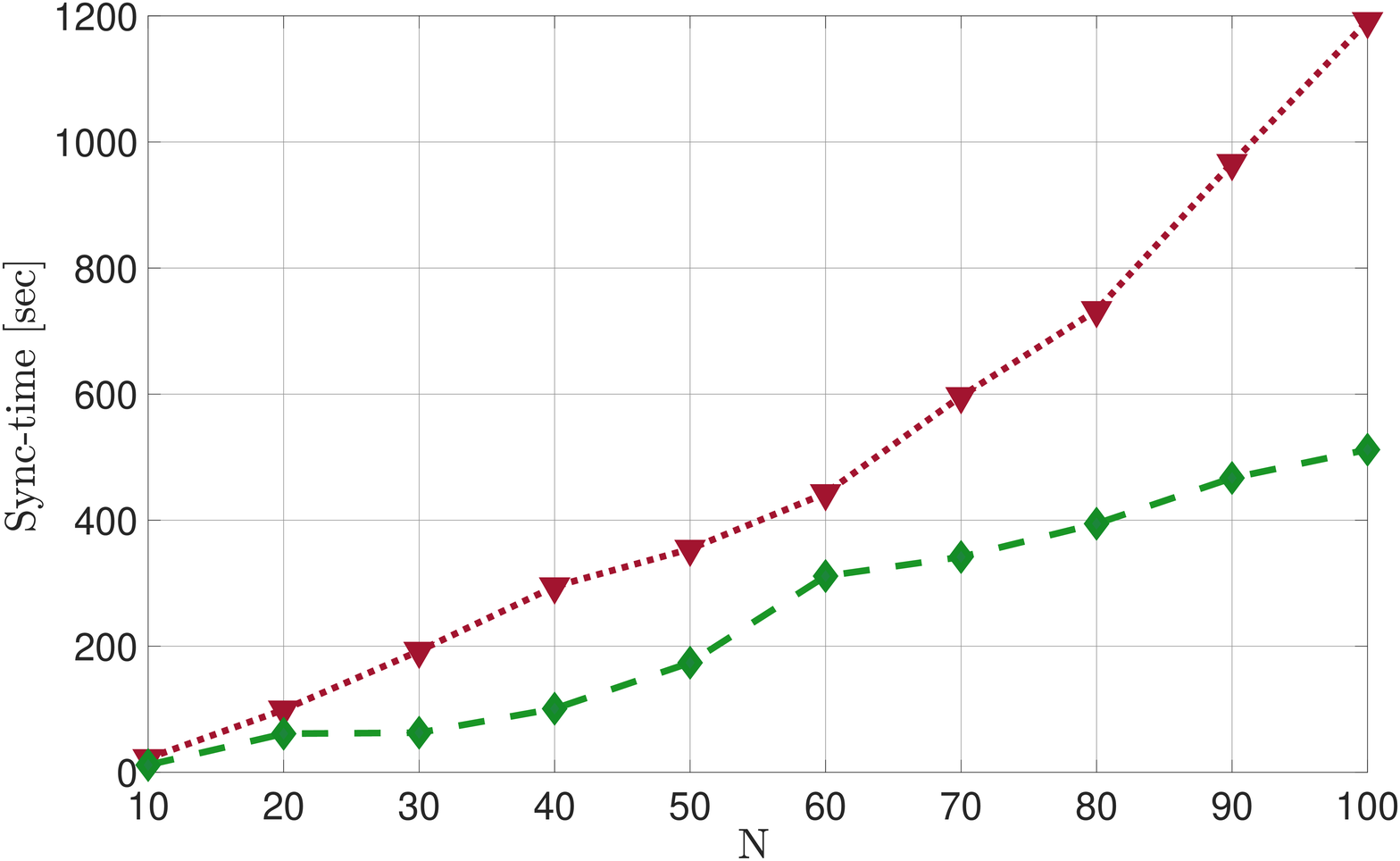}
    \hspace{-0.3cm} 
    \includegraphics[width=0.25\textwidth]{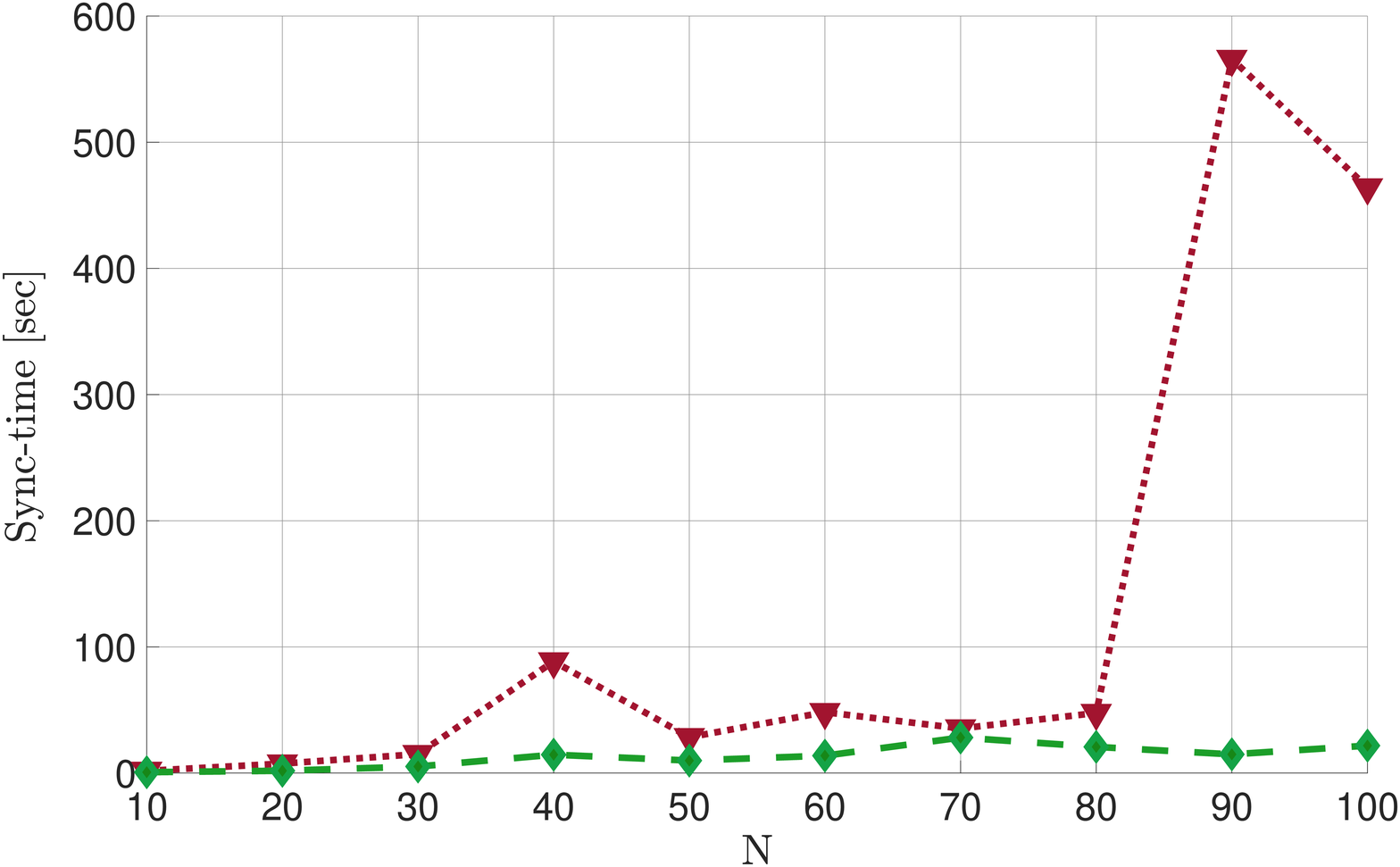}
    \caption{We plot the averages of  sync-times of our algorithm with binary jump map~\eqref{SHDS_graph} (depicted by the dashed, green curves) and the algorithm in~\cite{Klinglmayr} with piece-wise linear jump map~\eqref{eq:kling_jump_map} (depicted by the dotted, red curves) as functions of the number $N$ of agents. The network topologies, from left to right, are chosen to be complete-, path-, cycle-, and $5$-regular digraphs. For every such network topology and for every number $N$, the simulation results demonstrate that our algorithm synchronizes faster than the one proposed in~\cite{Klinglmayr}.}
    \label{fig:comparison}
\end{figure*}

By Lemma~\ref{lemmagraph2}, for any random solution $\pmb{\tau}_\omega$, there exist a hybrid time $(t^*_\omega, k^*_\omega)$, with $t^*_\omega\le T$, and a root $i^*$ of $\mathcal{G}$ such that $\pmb{\tau}_{\omega, i^*}(t^*_\omega, k^*_\omega) = 1$. 
Conditioning on the fact that $\pmb{\tau}_{\omega, i^*}(t^*_\omega, k^*_\omega) = 1$, we consider the event $\Omega_2(k^*_\omega, i^*)$. For convenience, we let  $t_\omega^{**}$ be the continuous-time instant corresponding to the $(k_\omega^* + L^*)$th jump. By Lemma \ref{lem:correspondence}, the number of jumps in a continuous-time period of $T$ is bounded below by 1. Then, for the discrete time to increase from $k^*_\omega$ to $k^*_\omega + L^*$, the continuous-time interval will increase by at most $L^*T$, i.e., we have that $ t_\omega^{**} - t_\omega^* \le L^* T$.
Next, by definition of the event $\Omega_2( k_\omega^*,i^*)$, the underlying digraphs between hybrid times $(t_\omega^*,k_\omega^*)$ and $(t_\omega^{**},k_\omega^*+L^*)$ are given by the synchronization string $\zeta$ (see Figure \ref{fig:result_stochastic} for an illustration). Thus, by the same arguments of  Theorem~\ref{thm:deterministic},   the random solution $\pmb{\tau}_\omega$ will reach synchronization before $(t_\omega^{**},k_\omega^*+L^*)$  provided that event $\Omega_2( k_\omega^*,i^*)$ is true. Since  $t_\omega^{*} \leq T$ and  $t_\omega^{**}-t_\omega^{*}\leq L^*T \leq \dep(\mathcal{G})\ell^* T$ and  since $\mathcal{A}_s$ is forward invariant, we have that $V(\pmb{\tau}_\omega(t_\omega, k_\omega)) = 0$, for all $t_\omega \ge T^*$. Thus, to establish~\eqref{eq:defineOmega4}, 
it now remains to show that the probability of the event $\Omega_2(j_\omega^*, i^*)$ is nonzero; by Lemma~\ref{lem:probability}, $\mathbb{P}(\Omega_2(j_\omega^*,  i^*)) = \eta$. Thus, the triplet $(p,r, \mathcal{G})$ is a sync-triplet.

\begin{figure}[t]
    \centering
\includegraphics[width=0.49\textwidth]{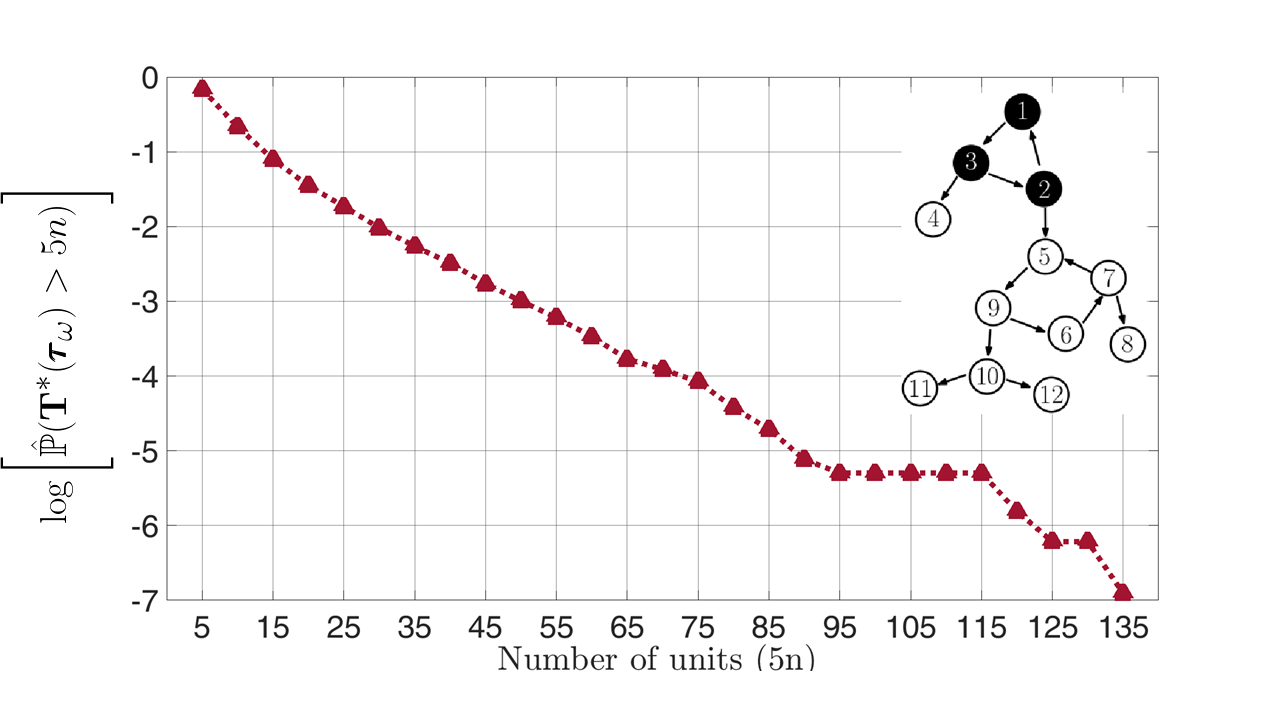} 
   \caption{Plot of $\log\bigl[\hat{\mathbb{P}}(\mathbf{T}^*(\pmb{\tau}_\omega)>5n)\bigr]$ versus the number of units $5n$ needed for a sample path to achieve synchronization. There are $1000$ random solutions simulated.}
   \label{fig:simulation1}
   \vspace{-0.1cm}
\end{figure}

Finally,  we show that~\eqref{stochasticbound} holds. First, by the Bayes rule, 
\begin{multline*}
\mathbb{P}\left 
(\pmb{T^*}(\pmb{\tau}_{\omega}) >  n T^*  
\right ) 
=  \mathbb{P}\left 
(\pmb{T^*}(\pmb{\tau}_{\omega}) > (n - 1) T^*  
\right )\hdots \\
~~~~~~~\hdots\times\mathbb{P}\left 
(\pmb{T^*}(\pmb{\tau}_{\omega}) > n T^*  
\, \vert \, \pmb{T^*}(\pmb{\tau}_{\omega}) > (n - 1) T^* \right ).
\end{multline*}
The conditional probability on the right hand side of the above expression can further be simplified as $\mathbb{P}(\pmb{T^*}(\pmb{\tau'}_{\omega'}) > T^* )$,  where $\pmb{\tau}'_{\omega'}$ is a new random solution with the initial condition $\tau'_{\omega'}(0,0)$ given by $\tau'_{\omega'}(0,0) = \tau_\omega((n - 1)T^*, k_\omega)$, for some $k_\omega$ and $\omega' := \omega_{k + 1} \omega_{k + 2} \cdots$. Note that by definition of $\Omega_3(\tau'_{\omega'}(0,0))$ and~\eqref{eq:defineOmega4}, we have that 
$$\mathbb{P}\left 
(\pmb{T^*}(\pmb{\tau'}_{\omega'}) > T^*
\right ) = 1- \mathbb{P}(\Omega_3(\tau'_{\omega'}(0,0))) < 1- \eta = \rho. $$
It then follows that 
$$\mathbb{P}\left 
(\pmb{T^*}(\pmb{\tau}_{\omega}) \ge  n T^*  
\right ) < \rho  \mathbb{P}\left 
(\pmb{T^*}(\pmb{\tau}_{\omega}) \ge  (n - 1) T^*  
\right ). $$
 The above recursive formula then implies that~\eqref{stochasticbound} holds. \hfill $\blacksquare$

\section{Simulation Results}
\label{sec:simulation}

In this section, we present numerical studies of the proposed algorithm \eqref{SHDS_graph}. We set $T=1$ and $p=0.5$. 

First, we verify the validity of Theorem~\ref{thm:main} and investigate the sync-time  $\pmb{T}^*(\pmb{\tau}_\omega)$ defined in \eqref{firsthitting}. 
For this purpose, we consider a rooted network of $N=12$ PCOs, as shown in Figure \ref{fig:simulation1}. Next, we  let the parameters $r_i$ be chosen uniformly randomly from $(0,1)^N$  and then  choose $1000$ random initial conditions uniformly  from $(0,1)^N$. For each initial condition, we simulate the SHDS \eqref{SHDS_graph} and let $(5(n-1),5n]$, for $n\ge 1$, be the interval that contains the sync-time. In Figure \ref{fig:simulation1}, we plot (in log scale) the empirical version of $\mathbb{P}(\pmb{T}^*(\pmb{\tau}_\omega)>5n)$  for different units $5n$, $n\ge 1$, i.e., we plot  $\hat{\mathbb{P}}(\pmb{T}^*(\pmb{\tau}_\omega)>5n):=1-\sum_{k=1}^n \frac{\text{Freq}(k)}{1000}$ where $\text{Freq}(k)$ is the total number of times that $\pmb{T}^*(\pmb{\tau}_\omega)$ belongs to $(5(n-1),5n]$.

Next, we compare the performance of our binary resetting algorithm \eqref{SHDS_graph} with the algorithm  considered in~\cite{Klinglmayr} (where the authors use a piece-wise linear jump map for numerical studies).   
  In the absence of delays, we reproduce their  piece-wise linear jump map
$H(z)$  below:
\begin{align}
\label{eq:kling_jump_map}
H(z)=\begin{cases}
h_1(z)=m_1z&0 \leq z\leq 0.5\\ h_2(z)=m_2z+1-m_2& 0.5 < z\leq 1
\end{cases}
\end{align}
where $m_1$ and $m_2$ are tuning parameters with $0<m_1\leq 0.5$ and $0<m_2\leq 0.5$. To be consistent with their algorithm, we let the parameters $r_i$ of our algorithm be $0.5$.  
The metric of performance is chosen to be the \textit{sync-time} \eqref{firsthitting}. Note that if one uses the algorithm in~\cite{Klinglmayr}, then reaching synchronization is only {\em asymptotic} with probability one. Thus, we relax the criterion of reaching synchronization such that the Lyapunov function $V$ defined in~\eqref{Lyapunov_function} only needs to satisfy $V(\pmb{\tau}_\omega) \leq 0.05$.  Correspondingly, we modify the sync-time $\pmb{T}^*(\pmb{\tau}_\omega)$ to be  $\pmb{T}_{0.05}^*(\pmb{\tau}_\omega):=\min_{t\ge 0}\{t: V(\pmb{\tau}_\omega)\leq 0.05 \}$.  

We first set $m_1=0.3261$ and $m_2=0.46$ as was done in the numerical studies of~\cite{Klinglmayr}. 
We run simulations for both algorithms for four different classes of network topologies: complete digraphs, path digraphs, cycle digraphs, and $5-$regular digraphs (see Section \ref{sec:Preliminaries} for definition). 
For each class of digraphs, we increase the number $N$ of agents from  $10$ to $100$, with the step of increment being $10$. Then, for each $N$, we generate 50 initial conditions uniformly randomly from $(0,1)^N$  used for both algorithms. In Figure \ref{fig:comparison}, we plot the averaged sync-time for comparison.

\begin{figure}[t]
    \centering
 \includegraphics[width=0.48\textwidth]{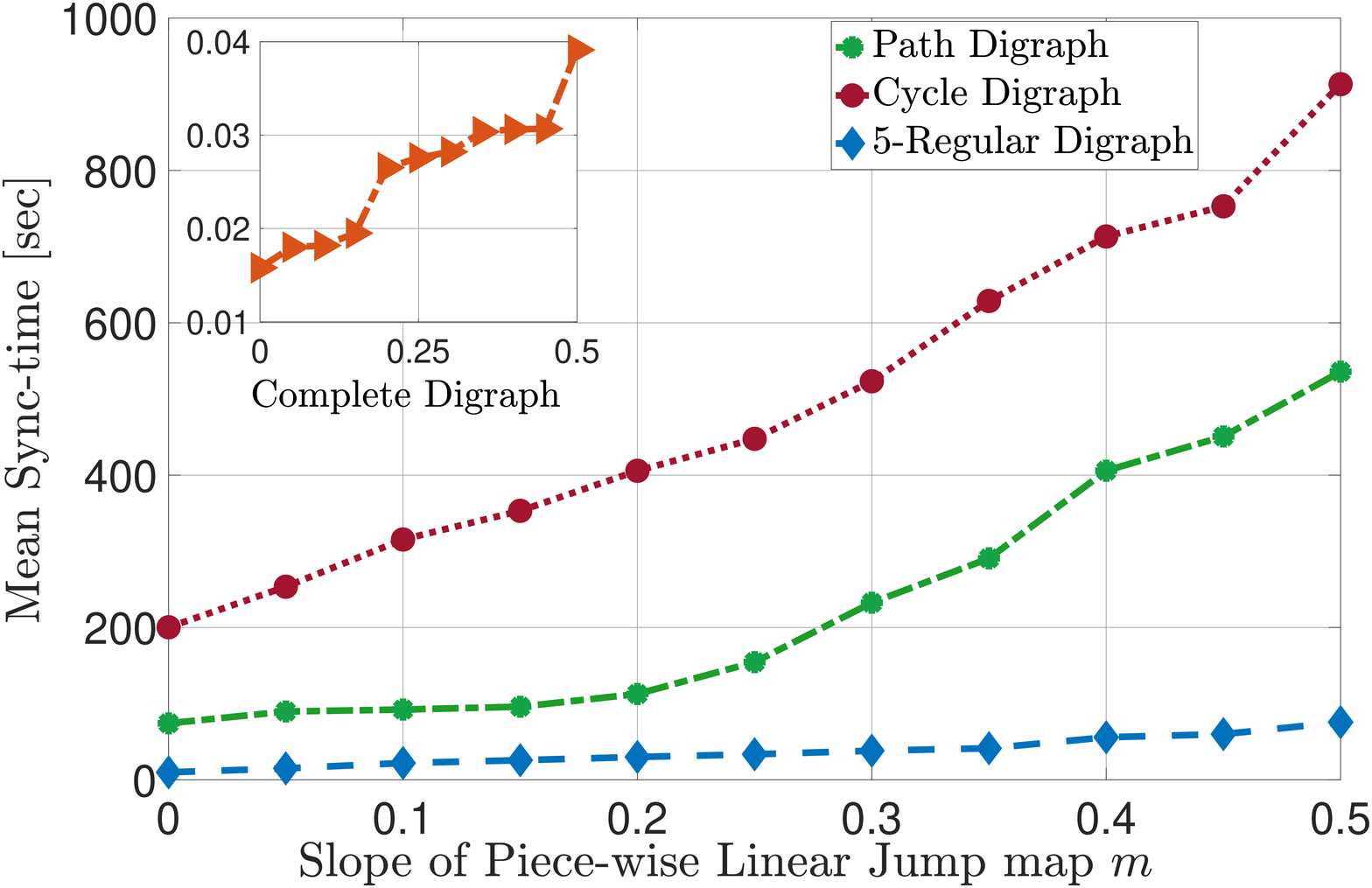}
   \caption{Averaged sync-times vs the slope $m_1=m_2=m$ of linear jump map of vertex-triggering algorithm in  \cite{Klinglmayr} for $N=50$.  }
   \label{fig:slope}
   \vspace{-0.1cm}
\end{figure}

Next, inspired by the use of piece-wise linear jump map in \cite{Klinglmayr}, we investigate via simulations how the slopes $m_1, m_2$ of the linear maps affect the sync-time. 
Note that the binary jump map can be viewed as an extremum case of the piece-wise linear map in a sense that the slopes  of the two linear functions in \eqref{eq:kling_jump_map} are $0$, i.e. $m_1=m_2=0$.   
Now, we set $m_1=m_2=:m$ and study the average sync-time as a function of~$m$. To this end, we fix $N=50$ vertices and consider again path-, cycle-, complete-, $5$-regular digraphs. We increase $m$ from $0$ to $0.5$ with the step of increment being $0.05$. For each digraph and for each $m$, we generate $50$ random initial conditions and run the simulations. We plot the averaged sync-time as a function of~$m$ in Figure \ref{fig:slope} for each digraph. It is observed that the averaged sync-time is the least when $m = 0$.

\section{Conclusion}
\label{sec:conclusion}
In this paper, we have presented a stochastic {\em binary, vertex-triggering} resetting algorithm by which networks of pulse-coupled oscillators can achieve global synchronization over rooted digraphs almost surely. The result is stated in Theorem~\ref{thm:main}. Its proof relies on the use of a hybrid-system machinery and the  analysis of the asymptotic behavior of a typical random solution of an associated stochastic hybrid dynamical system. Numerical studies 
have shown that our algorithm outperforms (in terms of the time needed for synchronization) an existing vertex-triggering algorithm over several different classes of information flow topologies.

\bibliographystyle{IEEEtran}      
\bibliography{biblio}  

\appendix

\subsection{Proof of Lemma \ref{forward_invariant}.}
\label{App:proof}

Let $x = (\tau,\lambda)$ be a maximal solution of the HDS. It suffices
 to show that the HDS $\mathcal{H}_\xi$ renders the set $\mathcal{A}_s$ strongly forward-invariant for the sub-state $\tau$.   Let  $\mathcal{A}_1:=\{\mu \mathbf{1}_N: \mu \in [0,1]\}$ and $\mathcal{A}_{2}:=\{0,1\}^N$. Then, the set $\mathcal{A}_s$ in \eqref{synchronization_set} can be written as the union of the disjoint subsets $\mathcal{A}_1\setminus \mathcal{A}_2$, $\mathcal{A}_2\setminus \mathcal{A}_1$ and $\mathcal{A}_1\cap \mathcal{A}_2$. Next, for every $(t,k)\in \dom(x)$ with $\tau(t,k) \in \mathcal{A}_s$, we show that $\tau(t,k)$ satisfies the following two conditions: First, if $\tau(t,k) \in C_\tau$, then for any $s\geq t$ with $(s,k)\in \dom (x)$ we have $\tau(s,k)\in \mathcal{A}_s$; Second, if $\tau(t,k) \in D_\tau$, then $\tau(t,k+1)\in \mathcal{A}_s$. 
To establish this, we consider the following three cases: (1) $\tau(t,k) \in \mathcal{A}_1\setminus \mathcal{A}_2$, (2) $\tau(t,k) \in \mathcal{A}_2\setminus \mathcal{A}_1$, and (3) $\tau(t,k) \in \mathcal{A}_1\cap \mathcal{A}_2$.
\vspace{.1cm}

\noindent
{\em Case (1).} Since $\mathcal{A}_1\setminus \mathcal{A}_2\subset C_\tau \setminus D_\tau$, $\tau(t,k) \in C_\tau \setminus D_\tau$.  Then, for $s \geq t$ with $(s,k)\in \dom (x)$,  sub-state $\tau$ flows as:
    \begin{equation}
     \label{eq:proof_flow} \tau(s,k)=\frac{1}{T}(s-t)\mathbf{1}_N+\tau(t,k).
    \end{equation} Because $\tau(t,k)=\mu \mathbf{1}_N$ for $\mu \in (0,1)$, we have that $\tau(s,k) \in \mathcal{A}_1\setminus \mathcal{A}_2.$ Hence,  $\tau(s,k) \in \mathcal{A}_s$.
 \vspace{.1cm}

\noindent   
{\em Case (2).} Since $\mathcal{A}_2\setminus \mathcal{A}_1\subset D_\tau$, $\tau(t,k) \in  D_\tau$. Then, the sub-state $\tau$ is updated as:
    \begin{equation}
     \label{eq:proof_jump}\tau(t,k+1) \in G_{\lambda(t,k)+1}(\tau(t,k)).
    \end{equation}
Because $\tau(t,k) \in D_\tau$, there exists at least one agent $i$ such that $\tau_i(t,k)=1$. If agent $i$ is the only one satisfying $\tau_i(t,k)=1$, then $\tau(t,k+1) = \mathbf{0}_N \in \mathcal{A}_1\cap \mathcal{A}_2$. Otherwise,  we have that $\tau(t,k+1) \in \mathcal{A}_2\setminus \mathcal{A}_1$. Hence,  $\tau(t,k+1) \in \mathcal{A}_s.$
\vspace{.1cm}

\noindent   
{\em Case (3).}  Since  $\mathcal{A}_1\cap\mathcal{A}_2= \{\mathbf{0}_N,\mathbf{1}_N  \}$, either $\tau(t,k) = \mathbf{0}_N$ or $\tau(t,k) = \mathbf{1}_N$. 
    We deal with the two sub-cases separately. 
    First, we assume that $\tau(t,k)=\mathbf{0}_N$. Then, for any $s \geq t$ with $(s,k)\in \dom (x)$, the sub-state $\tau$ evolves according to \eqref{eq:proof_flow}, which implies that $\tau(s,k)\in \mathcal{A}_1\setminus \mathcal{A}_2.$ Hence,  $\tau(s,k)\in \mathcal{A}_s$.
    Finally, we assume that $\tau(t,k)=\mathbf{1}_N$. Then, the sub-state $\tau$ evolves according to \eqref{eq:proof_jump}, which implies that $\tau(t,k+1)\in \mathcal{A}_2\setminus \mathcal{A}_1.$ Hence, for either of the two sub-cases, we have that $\tau(t,k+1)\in \mathcal{A}_s$. This concludes the proof. \hfill $\blacksquare$

\subsection{Hybrid Dynamical Systems}

    \label{App:HDS}
   Solutions of HDS \eqref{HDS} are parameterized by both continuous- and discrete-time indices $t\in\mathbb{R}_{\geq0}$ and $k\in\mathbb{Z}_{\geq0}$. A compact hybrid time domain is a subset of $\mathbb{R}_{\geq0}\times\mathbb{Z}_{\geq0}$ of the form $\cup_{k=0}^K([t_k,t_{k+1}]\times\{k\})$ for some $K\in\mathbb{Z}_{\geq0}$ and real numbers $0=t_0\leq t_1\leq\ldots\leq t_{K+1}$. A hybrid time domain is a set $E\subset\mathbb{R}_{\geq0}\times\mathbb{Z}_{\geq0}$ such that for each $T,K$, the set $E\cap ([0,T]\times\{0,1,2,\ldots,K\})$ is a compact hybrid time domain. A function $x:E\to\mathbb{R}^n$ is said to be a hybrid arc if $E$ is a hybrid time domain, and for each $k$ such that the interval $I_k=\{t\geq0:(t,k)\in\text{dom}(x)\}$ has non-empty interior the function $t\mapsto x(t,k)$ is locally absolutely continuous. A hybrid arc $x$ is said to be a solution to a HDS \eqref{HDS} satisfying the basic conditions of Definition \ref{definitionbasic1} if: (1) $x(0,0)\in C\cup D$. (2) If $(t_1,k), (t_2,k) \in \text{dom}(x)$ with $t_1<t_2$, then for almost every $t \in [t_1,t_2],$ $x(t,k) \in C$ and $\dot{x}(t,k) = f(x(t,k))$. (3) If $(t,k),(t,k+1) \in \text{dom}(x)$, then $x(t,k) \in D$ and $x(t,k+1) \in G(x(t,k))$.

\subsection{Stochastic Hybrid Dynamical Systems}
 \label{App:SHDS}
 
 \counterwithin{thm}{subsection}

   Random solutions to SHDS \eqref{SHDS1} are functions of $\omega\in\Omega$ denoted ${\bf x}(\omega)$, such that: 1) $\omega\mapsto {\bf x}(\omega)$ has measurability properties that are adapted to the minimal filtration of ${\bf v}$; 2) for each $\omega\in\Omega$ the sample path ${\bf x(\omega)}$ is a standard solution to the HDS \eqref{HDS} with the appropriate causal dependence on the random input ${\bf v(\omega)}$ through the jumps. To formally define these mappings, for $\ell\in\mathbb{Z}_{\geq1}$, let $\mathcal{F}_\ell$ denote the collection of sets $\{\omega\in\Omega:({\bf v}_1(\omega),{\bf v}_2(\omega),\ldots,{\bf {v}}_\ell(\omega))\in F\}$, $F\in\mathbf{B}(\mathbb{R}^m)^\ell)$, which are the sub-$\sigma$-fields of $\mathcal{F}$ that form the minimal filtration of ${\bf v}=\{{\bf v}_\ell \}_{\ell=1}^{\infty}$, which is the smallest $\sigma$-algebra on $(\Omega,\mathcal{F})$ that contains the pre-images of $\mathbf{B}(\mathbb{R}^m)$-measurable subsets on $\mathbb{R}^m$ for times up to $\ell$. A stochastic hybrid arc is a mapping ${\bf x}$ from $\Omega$ to the set of hybrid arcs, such that the set-valued mapping from $\Omega$ to $\mathbb{R}^{n+2}$, given by  $\omega\mapsto \text{graph}({\bf x}(\omega)):=\big\{(t,k,z):\tilde{x}={\bf x}(\omega), (t,k)\in\text{dom}(\tilde{x}),z=\tilde{x}(t,k)\big\}$, is $\mathcal{F}$-measurable with closed-values. Let $\text{graph}({\bf x}(\omega))_{\leq \ell}:=\text{graph}({\bf x} (\omega))\cap (\mathbb{R}_{\geq0}\times\{0,1,\ldots,\ell\}\times\mathbb{R}^n)$. An $\{\mathcal{F}_\ell\}_{\ell=0}^{\infty}$ adapted stochastic hybrid arc is a stochastic hybrid arc ${\bf x}$ such that the mapping $\omega\mapsto \text{graph}({\bf x}(\omega))_{\leq \ell}$ is $\mathcal{F}_\ell$ measurable for each $\ell \in\mathbb{N}$. An adapted stochastic hybrid arc ${\bf x}(\omega)$, or simply $\mathbf{x}_\omega$,  is a solution to SHDS \eqref{SHDS1}, satisfying the basic conditions of Definition \ref{definitionbasic1}, starting from $x_0$ denoted ${\bf x}_\omega\in \mathcal{S}_r(x_0)$ if: (1) $\mathbf{x}_{\omega}(0,0)=x_0$; (2) if $(t_1,k),(t_2,k)\in\text{dom}(\mathbf{x}_{\omega})$ with $t_1<t_2$, then for all $t\in[t_1,t_2]$, $\mathbf{x}_{\omega}(t,k)\in C$ and $\dot{\mathbf{x}}_{\omega}(t,k)= f(\mathbf{x}_{\omega}(t,k))$; (3) if $(t,k),(t,k+1)\in\text{dom}(\mathbf{x}_{\omega})$, then $\mathbf{x}_{\omega}(t,k)\in D$ and $\mathbf{x}_{\omega}(t,k+1)\in G(\mathbf{x}_{\omega}(t,k),\mathbf{v}_{k+1}(\omega))$. A random solution ${\bf x}_\omega$ is said to be: a) almost surely {\em non-trivial} if its hybrid time domain contains at least two points almost surely; b) almost surely  {\em  complete} if for almost every sample path $\omega\in \Omega$ the hybrid arc ${\bf x}_\omega$ has an unbounded time domain; and almost surely eventually discrete if for almost every sample path $\omega\in \Omega$ the hybrid arc ${\bf x}_\omega$ is eventually discrete. A continuous function $V:\mathbb{R}^n\to\mathbb{R}_{\geq0}$ is a Lyapunov function relative to a compact set $\mathcal{A}\subset\mathbb{R}^n$ for the SHDS \eqref{SHDS1} if $V({\bf x}_\omega)=0\iff {\bf x}_\omega \in\mathcal{A}$, $V$ is radially unbounded with respect to set $\mathcal{A}$, non-increasing during flows, and $\int_{R^m}\max_{g\in G({\bf x}_\omega,v)} V(g)\mu(dv)\leq V({\bf x}_\omega),~~\forall~{\bf x}_\omega\in D$. The following stochastic hybrid invariance principle \cite[Thm. 8]{rec_principle} is instrumental for our analysis of Theorem \ref{thm:main}.

\vspace{0.2cm}
\begin{thm}\label{S_rec}
Let $V$ be a Lyapunov function relative to a compact set $\mathcal{A} \subset \mathbb{R}^n$ for the SHDS system $\mathcal{H}$. Then, $\mathcal{A}$  is UGASp if and only if there does not exist an almost surely complete solution ${\bf x}_\omega$ that remains in a non-zero level set of the Lyapunov function almost surely. 
\end{thm}
  
\begin{IEEEbiography}[{\includegraphics[width=1in,height=1.25in,clip,keepaspectratio]{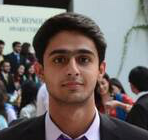}}]{Umar Javed}  received in 2017 his B.S. degree in Electrical Engineering with minors in Computer Science (AI/ML) and Psychology 
from the Lahore University of Management Sciences, Pakistan. Currently, he is a PhD candidate in the Department of Electrical, Computer, and Energy Engineering at the University of Colorado, Boulder, where he completed his MS in Electrical Engineering in 2019. 
\end{IEEEbiography}

\begin{IEEEbiography}[{\includegraphics[width=1in,height=1.25in,clip,keepaspectratio]{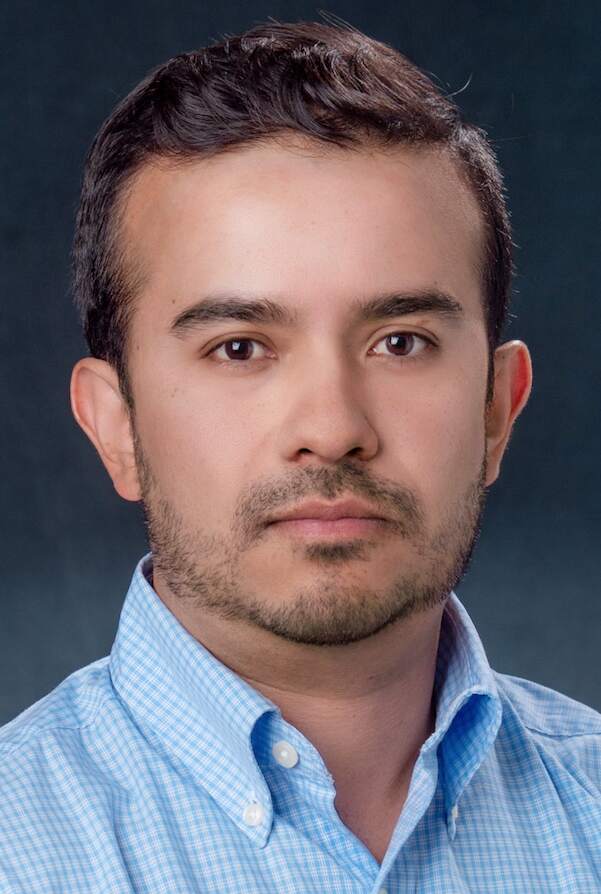}}]{Jorge I. Poveda}  is an Assistant Professor in the Department of Electrical, Computer, and Energy Engineering at the University of Colorado, Boulder. He received the M.Sc. and Ph.D. degrees in Electrical and Computer Engineering from the University of California at Santa Barbara in 2016 and 2018, respectively, where he was awarded the CCDC Outstanding Scholar Fellowship and the Best Ph.D. Thesis Award. Before joining CU Boulder in 2019, he was a Postdoctoral Fellow at Harvard University. Dr. Poveda is an awardee of the 2022 Air Force's Young Investigator Research Program (YIP), a recipient of the NSF CRII (2020) and Career (2022) Awards, as well as the campus-wide RIO Faculty Fellowship at CU Boulder in 2022.
\end{IEEEbiography}

\begin{IEEEbiography}[{\includegraphics[width=1in,height=1.25in,clip,keepaspectratio]{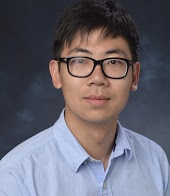}}]{Xudong Chen}  is an Assistant Professor in the Department of Electrical, Computer and Energy Engineering at the University of Colorado, Boulder. Before that, he was a postdoctoral fellow in the Coordinated Science Laboratory at the University of Illinois, Urbana-Champaign. He obtained the B.S. degree in Electronics Engineering from Tsinghua University, Beijing, China, in 2009, and the Ph.D. degree in Electrical Engineering from Harvard University, Cambridge, Massachusetts, in 2014. Dr. Chen is an awardee of the 2020 Air Force's Young Investigator Research Program (YIP), a recipient of the 2021 NSF Career Award, and the recipient of the 2021 Donald P. Eckman Award.  
\end{IEEEbiography}

\end{document}